\documentclass[10pt,twocolumn]{IEEEtran} 	

\usepackage{fancyhdr, epsfig, epsf, amsthm, amsmath, amssymb, amsfonts, subfigure, color, graphicx}
\usepackage{threeparttable}
\usepackage[noadjust]{cite}
\usepackage{dsfont}
\usepackage{enumerate}
\usepackage{comment}

\newtheorem{theorem}{Theorem}

\newtheorem{definition}[theorem]{Definition}

\newtheorem{lemma}{Lemma}
\newtheorem{corollary}{Corollary}

\newtheorem{proposition}{Proposition}
\newtheorem{remark}{Remark}

\def\qed{$\blacksquare$}

\def\endproof{\hfill \qed}

\def\E{\mathsf{E}}
\def\SINR{\mathsf{SINR}}
\def\SNR{\mathsf{SNR}}
\def\SIR{\mathsf{SIR}}

\def\l{\left}
\def\r{\right}
\def\({\left(}
\def\){\right)}
\def\lb{\left\{}
\def\rb{\right\}}
\def\[{\left[}
\def\]{\right]}

\newcommand{\nn}{\nonumber}

\includecomment{comment}	

\def\lambdahM{{\hat{\lambda}_m}{}}

\def\lambdahMu{{\hat{\lambda}_\mu}{}}
\def\lambdaM{\lambda_{m}}
\def\lambdaMu{\lambda_\mu}
\def\lambdaU{\lambda_{u}}
\def\gammaM{\gamma_{m}}
\def\gammaMu{\gamma_\mu}
\def\gammaMUL{\gamma_{m.u}}
\def\gammaMDL{\gamma_{m.d}}
\def\gammaMuUL{\gamma_{\mu.u}}
\def\gammaMuDL{\gamma_{\mu.d}}
\def\WM{W_m{}}
\def\WMu{W_\mu{}}
\def\WMUL{W_{m.u}{}}

\def\RL{R_{L}{}} 

\def\pLt{p_{L}^{(t)}}
\def\pL{p_{L}{}}
\def\rhoMu{\rho_\mu}
\def\rhoMut{\rhoMu^{(t)}}

\def\rhoM{\rho_{m}}
\def\rhoMt{\rho_{m}^{(t)}}

\def\alphaM{\alpha_m}
\def\alphaMu{\alpha_\mu}
\def\RDL{\mathcal{R}_d{}}
\def\RUL{\mathcal{R}_u{}}

\def\PAPR{\text{PAPR}}
\def\PhiActU{\Phi_{\dot{u}}}
\def\PhiActM{\Phi_{\dot{m}}}
\def\PhiActMu{\Phi_{\dot{\mu}}}

\def\Pdac{P_{\text{dac}}}
\def\cdac{c_{\text{dac}}}
\def\lambdahS{\hat{\lambda}_{S}}
\def\lambdahU{\hat{\lambda}_{U}}

\def\papertitle{Tractable Resource Management with Uplink Decoupled Millimeter-Wave Overlay in Ultra-Dense Cellular Networks}

\IEEEoverridecommandlockouts 

\begin{document}

\title{ \fontsize{21}{21}\selectfont \papertitle}

\author{Jihong Park, Seong-Lyun Kim, and Jens Zander
\thanks{This work was supported by the National Research Foundation of Korea (NRF-2014R1A2A1A11053234) and the Communication Policy Research Center (CPRC) support program supervised by the Institute for Information and Communications Technology Promotion (IITP-2015-H8201-15-1003), funded by the Ministry of Science, ICT and Future Planning of Korea.}

\thanks{J. Park and S.-L. Kim are with School of Electrical and Electronic Engineering, Yonsei University, Seoul 120-749, Korea, E-mail: \{jhpark.james, slkim\}@ramo.yonsei.ac.kr.  }
\thanks{J. Zander is with the Center for Wireless Systems (wireless@kth), Department of Communication Systems, School of Information and Communication Technology, KTH Royal Institute of Technology, Stockholm 164 40, Sweden, E-mail: jenz@kth.se.}}

\maketitle

\begin{abstract} 
The forthcoming 5G cellular network is expected to overlay millimeter-wave (mmW) transmissions with the incumbent micro-wave ($\mu$W) architecture. The overall mm-$\mu$W resource management should therefore harmonize with each other. This paper aims at maximizing the overall downlink (DL) rate with a minimum uplink (UL) rate constraint, and concludes: mmW tends to focus more on DL transmissions while $\mu$W has high priority for complementing UL, under time-division duplex (TDD) mmW operations. Such UL dedication of $\mu$W results from the limited use of mmW UL bandwidth due to excessive power consumption and/or high peak-to-average power ratio (PAPR) at mobile users. To further relieve this UL bottleneck, we propose mmW UL decoupling that allows each legacy $\mu$W base station (BS) to receive mmW signals. Its impact on mm-$\mu$W resource management is provided in a tractable way by virtue of a novel closed-form mm-$\mu$W spectral efficiency (SE) derivation. In an ultra-dense cellular network (UDN), our derivation verifies mmW (or $\mu$W) SE is a logarithmic function of BS-to-user density ratio. This strikingly simple yet practically valid analysis is enabled by exploiting stochastic geometry in conjunction with real three dimensional (3D) building blockage statistics in Seoul, Korea.
\end{abstract}

\section{Introduction}
Spectrum bandwidth increase has been playing a key role to improve data rate in cellular history. The upcoming 5G cellular networks aiming at the 1,000-fold rate improvement, however, can no longer count solely on the existing micro-wave ($\mu$W) spectrum bandwidth. The exploration to resolve this spectrum scarcity has recently reached the use of millimeter-wave (mmW) spectrum whose bandwidth is hundreds times larger than $\mu$W bandwidth \cite{SamsungmmWave:11, SamsungSarnoff11, SamsungGC:13, Ericsson5G:13, Rappaport5G:13, Rappaport:14, Andrews5G:14}. It leads to the dual radio access technology (RAT) operations in 5G where mmW transmissions are overlaid on top of the legacy $\mu$W cellular architecture \cite{SamsungmmWave:11, SamsungSarnoff11, Ericsson5G:13}.

To reach the highest capacities for 5G, interest has lately been turning toward ultra-dense BS deployment \cite{Ericsson5G:13,Andrews5G:14,Holistic13, Zander13,Qualcomm:14}. It is an effective way to enhance $\mu$W cellular capacity in place of bandwidth increase under the $\mu$W spectrum depletion. Besides, this approach fits nicely into mmW transmissions. Deploying more BSs assists in assuring line-of-sight (LOS) conditions for mmW transmissions. This thereby mitigates severe distance attenuation due to physical blockages, which is a major drawback of utilizing mmW spectrum \cite{mmWUDN:15}. In this respect, paving the way for  a mm-$\mu$W ultra-dense cellular network (UDN) is of our prime concern.

The critical challenge to such a network operation is the disjunction between downlink (DL) and uplink (UL) rates. It results from employing extremely wide mmW bandwidth at transmissions. Enabling the mmW transmissions brings about low energy efficiency \cite{SamsungmmWave:11, SamsungSarnoff11}, captured by high peak-to-average-power-ratio (PAPR) at power amplifiers and excessive power dissipation at digital-to-analog converters (DACs). Unlike BSs having sufficient power supply, mobile users may therefore not be able to exploit the entire mmW bandwidth for transmissions, in spite of their having exclusive access to BSs allowed by a minuscule number of UDN per-cell users. This leads to significant DL/UL rate discrepancy, which may impinge on user experiences during DL/UL symmetric services such as high-definition video calling. Furthermore, it may hold back improving DL rate when UL rate cannot cope with the required control signals for DL.

In view of tackling this bottleneck, we focus on the DL/UL resource management in a mm-$\mu$W network (see Fig. 1-a), and jointly adjust their allocations so as to maximize the overall mm-$\mu$W DL rate while guaranteeing a minimum UL/DL rate ratio (Fig. 1-b). For given mm-$\mu$W bandwidths, this paper considers the question: \emph{how much portion of mm-$\mu$W resources should be allocated to UL transmissions as mmW BS deployment proliferates}. When deploying more mmW BSs, it answers that $\mu$W increases UL allocation, followed by increasing the UL allocation of mmW after reaching the $\mu$W's dedication to UL. This sequential UL allocations lead to DL dedicated mmW spectrum during the early phase of mmW BS deployment, and UL dedicated $\mu$W spectrum when mmW BS deployment proliferates (see Propositions 4 and 6 in Section IV for further details). 

To derive such resource management, it requires the mm-$\mu$W spectral efficiency (SE) calculation. In this paper, we derive the closed-form SEs in a mm-$\mu$W UDN by using stochastic geometry. In addition to its providing tractability in mm-$\mu$W resource management, the closed-form result reveals the following UDN characteristics: (i) UDN SE is a \emph{logarithmic function of BS-to-user density ratio}; and (ii) a UDN is \emph{weak interference-limited} where its behaviors toward BS densification and blockages are analogous to the trends in a noise-limited regime.

Motivated by the fact UDN SE increases along with BS densification, we propose a novel \emph{mmW UL decoupling} technique that further alleviates the UL rate bottleneck. It allows legacy $\mu$W BSs to receive mmW UL signals (see user 2 in Fig.~1-a). In so doing, not only mmW resource is spatially reused but also mmW received signal power is increased, compensating the limited use of mmW UL bandwidth. The corresponding mm-$\mu$W UDN resource management and overall DL rate are numerically validated under a three dimensional (3D) mmW blockage model combined with the actual building statistics in Seoul, Korea.

\begin{figure*} \label{Fig:Network}
	\centering
	\subfigure[mm-$\mu$W network operations]{\includegraphics[width=10cm]{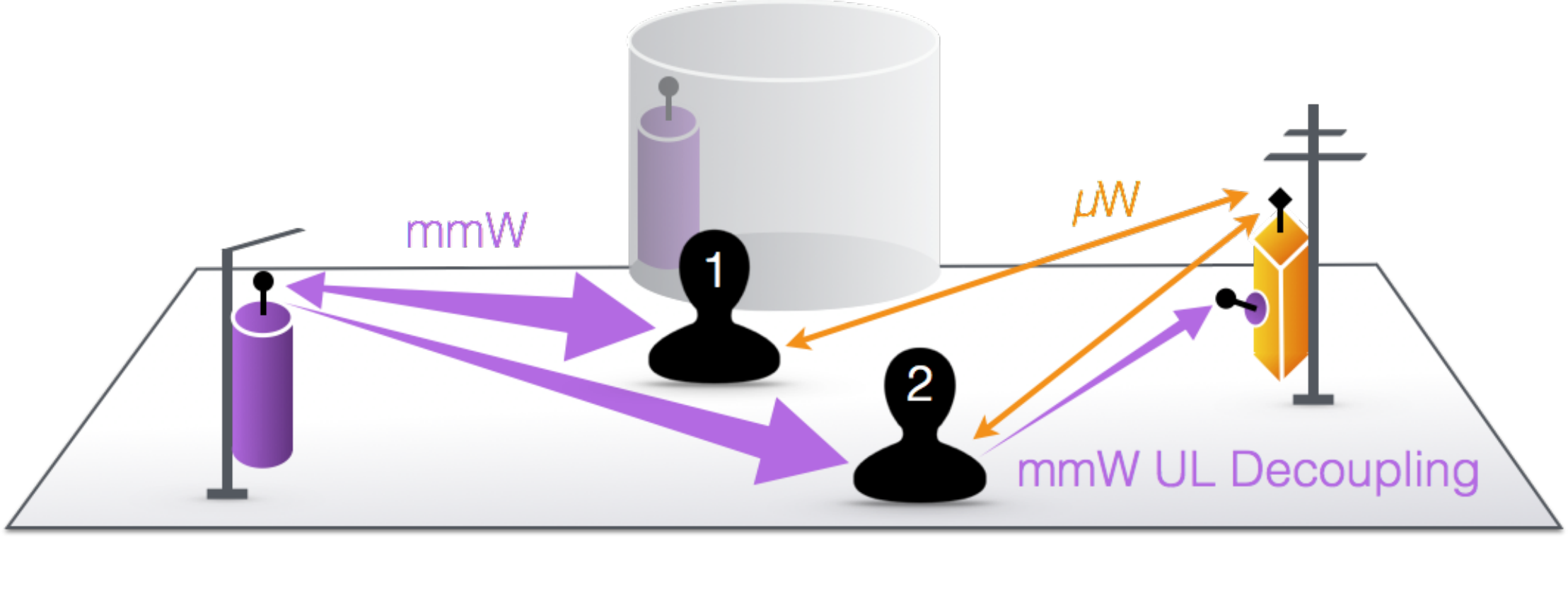}}
	\subfigure[mm-$\mu$W DL/UL resource management]{\includegraphics[width=6cm]{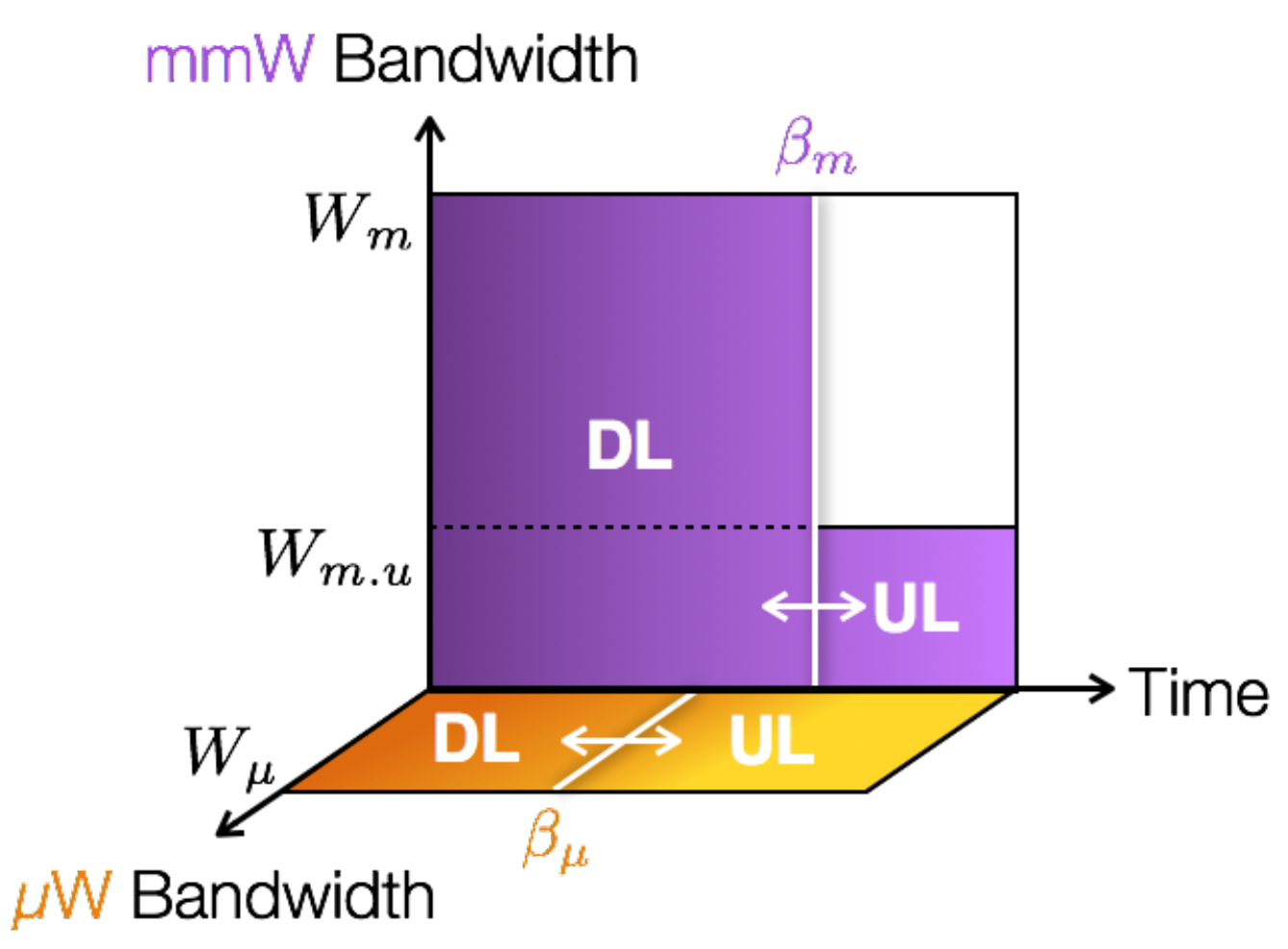}}
	\caption{Illustrations of mm-$\mu$W network operations and its DL/UL resource management under mmW non-penetrable building blockages: \textbf{Associations}. Each user has a pair of dual connections for mm-$\mu$W receptions/transmissions based on the strongest received power association rule. By the aid of \emph{mmW UL decoupling}, mmW UL users are also able to associate with $\mu$W BSs if they provide stronger UL signals than mmW BSs (compare user 2 with user 1); \textbf{DL/UL Rate Asymmetry}. mmW UL bandwidth is limited by PAPR outage and/or DAC power dissipation at mobiles, i.e. $\WMUL < \WM$, and thus its corresponding UL rate is smaller than DL rate; \textbf{DL/UL Allocations}. Given TDD operations, mm-$\mu$W UL allocation ratios are determined via adjusting $\beta_m$ and $\beta_\mu$.} \label{Fig:Network}
\end{figure*}

\subsection{Related Works}
Utilizing mmW spectrum in 5G cellular networks has recently been viewed as of paramount importance in both industry \cite{SamsungmmWave:11, SamsungGC:13, Ericsson5G:13} and academia \cite{Rappaport:14,Rappaport5G:13,Andrews5G:14, Heath:13,Heath:14}. Its point-to-point-wise cellular network applicability was validated by indoor/outdoor channel modeling \cite{Rappaport:14,Rappaport5G:13} as well as by laboratory demonstration \cite{SamsungGC:13}. For its network-wide deployment, the authors \cite{Heath:13,Heath:14} analyzed the mmW indoor/outdoor cellular network from the coverage probability and average rate perspectives. On the basis of these preceding efforts, this study considers both mmW and the incumbent $\mu$W coexisting heterogeneous networks that are highly promising phased deployment scenarios toward 5G \cite{Ericsson5G:13,Andrews5G:14}. Starting from this dual-RAT structure, we focus on a higher layer design guideline how to jointly manage the mm-$\mu$W radio resources for DL/UL transmissions.

When it comes to the UDN impact, previous investigations paid attention to its complementing the blockage-vulnerable drawback of mmW \cite{mmWUDN:15} and reducing interference \cite{Yu2011, SLeeKHuang12, Alexiou13, JHPark:14,JHParkAPWCS:15,JHParkGC:15,TowardUDN:15}. The decrease in UDN interference follows from the turned-off BSs having no serving users according to the Third Generation Partnership Project (3GPP) Release 12 specifications \cite{NCT2013}. This article embodies both ultra-densification effects on SE improvement, and provides closed-form mm-$\mu$W UDN SEs. 


Note that our preliminary research about mm-$\mu$W UDN \cite{JHParkGC:15, JHParkAPWCS:15} in part includes similar arguments to this article but under different scenarios. Specifially, an indoor/outdoor modeling is considered in \cite{JHParkAPWCS:15}, and a 3D blockage model with actual building geography is incorporated in \cite{JHParkGC:15}. Both works are based on frequency-division duplex (FDD) mmW operations unlike the time-division duplex (TDD) mmW operations in this article. From an analytic perspective, the closed-form $\mu$W UDN SE derivation was first introduced in \cite{JHPark:14} with its economic impact, based on a lower bound approximation. This article provides the upper bound for completeness, and also derives mmW UDN SE extending the analytic applicability. 

In addition, it is worth mentioning that this study is on the basis of our UDN definition where the number of BSs exceeds the number of users (see Definition 1) as in \cite{JHPark:14, JHParkAPWCS:15, JHParkGC:15,TowardUDN:15,EricssonVTC:15,LOSNLOS:15}. Therefore, the UDN SE arguments in this article may not be in line with other opinions, especially regarding the near-field effect that may degrade SE under a scenario neglecting the turned-off BS impact \cite{3GPPMultSlope:15,MultSlope:14,AndrewsUDN:15}. Incorporating the near-field effect with our UDN definition is deferred to the partially loaded scenario in \cite{LOSNLOS:15}.

\subsection{Contributions and Organizations}
The main contributions of this paper are listed as follows.
\begin{itemize}
\item \emph{Closed-form mm-$\mu$W UDN SEs} are derived, enabling tractable mm-$\mu$W resource management. The representations of both DL/UL mmW (or $\mu$W) UDN SEs are identically a logarithmic function of the BS density normalized by user density (see Theorems 1 and 2 in Section \ref{Sect:SEresult}).

\item The mm-$\mu$W UL resource allocation strategy is provided in order to maximize DL rate with a minimum UL/DL rate ratio constraint as follows: along with the mmW BS densification, \emph{$\mu$W UL allocation increases, and mmW UL allocation grows only after using up the entire $\mu$W bandwidth}. This in the end leads to the \emph{UL dedicated $\mu$W spectrum} that contradicts with the traditional $\mu$W cellular resource management trend (Propositions 4 in Section \ref{Sect:Opt_WandDens2}).

\item To further rein back DL/UL rate asymmetry, \emph{mmW UL decoupling} is proposed. The technique allows $\mu$W BSs to receive mmW UL siganls. This thereby increases mmW UL rate, resulting in up to $1.33$ times overall DL rate improvement in given scenarios (Propositions 6--7 in Section \ref{Sect:Opt_WandDens3}).

\item The mm-$\mu$W DL/UL resource allocations are numerically estimated on the basis of a \emph{realistic 3D building blockage model}. The calculation incorporates the actual building locations, sizes, and heights in Seoul, Korea (Table I in Section \ref{Sect:3DBlockage}).
\end{itemize}

The rest of the paper is organized as follows. The mm-$\mu$W network environment and its operations are specified in Section \ref{Sect:Sys}. The closed-form mm-$\mu$W UDN SEs are derived in Section \ref{Sect:SE}. The results are utilized for providing tractable mm-$\mu$W resource management as well as for emphasizing the mmW UL decoupling impact in Section \ref{Sect:RscMngmt}. Numerical evaluation under a real data based 3D building blockage model is presented in Section \ref{Sect:Numerical} followed by concluding remarks in Section \ref{Sect:Conclusion}. The proofs of lemmas and propositions are provided respectively in Appendices I and II.

\section{System Model} \label{Sect:Sys}
\subsection{Millimeter-Wave Overlaid Cellular Network} \label{Sect:Network}
A mm-$\mu$W network consists of mmW and $\mu$W BSs. The locations of mmW BSs follow two-dimensional (2D) homogeneous Poisson point process (PPP) $\Phi_{m}$ with density $\lambdaM$ \cite{StoyanBook:StochasticGeometry:1995}. Similarly, the $\mu$W BS coordinates follow a homogeneous PPP $\Phi_\mu$ with density $\lambdaMu$, independent of $\Phi_\text{m}$. Mobile user coordinates independently follow a homogeneous PPP $\Phi_\text{u}$ with density $\lambdaU$. Without loss of generality, $\Phi_\text{u}$ denotes either DL or UL users with the same density. For simplicity, we hereafter only focus on outdoor users whose coordinates do not overlap with building blockages. The blockage modeling is specified in Section \ref{Sect:SysBlockage}. 

As Fig.~\ref{Fig:Network} visualizes, users simultaneously associate with mmW and $\mu$W BSs via different antennas for each as in macro-and-small-cell dual connectivity \cite{Rel12Beyond:13}. The associations of users for both mmW and $\mu$W are based on the maximum received signal powers. Control signals are communicated via $\mu$W so as to guarantee the connections regardless of blockages. The control signal traffic is assumed to be negligible when calculating data rates since the volume is much less than that of data transmissions.

Regarding multiple associations of users at a BS, the BS selects a single active user per unit time slot according to a uniformly random scheduler. The selected active users are denoted as $\PhiActU$. The mmW (or $\mu$W) BSs having at least a single active user are represented by $\Phi_{\dot{m}}$ (or $\Phi_{\dot{\mu}}$). Note that $\PhiActU, \PhiActM$, and $\PhiActMu$ are non-homogeneous PPPs due to their location dependent active user selections; for instance, a user having more neighboring BSs (or smaller Voronoi cell size of the associated BS) is selected with higher probability, incurring location dependency. The mm-$\mu$W BSs having no associated users are turned-off.

\subsection{Blockage Model} \label{Sect:SysBlockage}
Consider building blockages that cannot be penetrated by mmW signals. Every single mmW signal reception should therefore ensure the line-of-sight (LOS) condition between its transmitter-receiver pair. According to the average LOS model \cite{Heath:13}, assume that LOS is guaranteed when a given transmitter-receiver distance $r$ is within an LOS distance $\RL$ that is determined by the site geography of a network. The LOS indicator function $\mathds{1}_{\RL}(r)$ represents such a model, which returns unity if LOS exists, i.e. $r \leq \RL$; otherwise zero. Note that mmW communication via non-LOS links is feasible in practice through reflections as pointed out in \cite{Rappaport:14,Heath:14,Heath:13}. Nevertheless, its impact on the the DL/UL rate asymmetry is negligible since this increases both DL and UL mmW rates, thus neglected in this paper. Analyzing the effect of mmW reflections is expected to be viable via utilizing multiple-slope path loss models \cite{LOSNLOS:15, MultSlope:14}, deferred to future work.

The value of $\RL$ with the 2D blockage model \cite{Heath:13,Heath:14,Singh:14,Kulkarni:14} is directly computed by utilizing building perimeters, areas, and coverage from actual geographical data. For the 3D blockage model, the calculation of $\RL$ additionally incorporates building heights, to be further elaborated in Section \ref{Sect:3DBlockage}. Throughout this paper, we by default consider the 3D blockage model, but apply a 2D channel model that provides more tractability. In contrast to mmW signals, $\mu$W signals are not affected by blockages thanks to their high diffraction and penetration characteristics.

\subsection{Channel Model}
For $\mu$W DL and UL, a signal is transmitted from a single omnidirectional antenna with the fixed powers respectively set as $P_{m.d}$ and $P_{m.u}$. It is worth noting that UL power control is neglected, which can be practically feasible under UDNs. Cell sizes in a UDN are extremely small, and therefore UL power control for cell edge users such as channel inversion \cite{SinghDecouple:14} loses its merit. The UL power control of course increases energy efficiency even under UDNs, albeit out of the main interest in this work.

The transmitted $\mu$W signal experiences path loss attenuation with the exponent $\alphaMu >2$ as well as Rayleigh fading with unity mean, i.e. channel fading power $g \sim \exp(1)$. When interference is treated as noise, by the aid of Slyvnyak's theorem \cite{StoyanBook:StochasticGeometry:1995}, DL/UL $\SINR$ at a typical DL user and at an UL Bs is then respectively represented as:

\vspace{-5pt}\small\begin{align}
\SINR_{\mu.d} &:= \frac{P_{\mu.d}g r^{-\alpha_\mu} }{\sum_{i\in \PhiActMu \text{(or $\PhiActU$)}} P_{\mu.d}g_i {r_i}^{-\alpha_\mu}+ \sigma^2} \\
\SINR_{\mu.u} &:= \frac{P_{\mu.u}g r^{-\alpha_\mu} }{\sum_{i\in \PhiActU \text{(or $\PhiActMu$})} P_{\mu.u}g_i {r_i}^{-\alpha_\mu}+ \sigma^2}.
\end{align}\normalsize

For mmW DL and UL, a signal is directionally transmitted from $N$ number of antennas respectively with power $P_{m.d}$ and $P_{m.u}$. Its beam pattern experienced at a receiver follows a simple model in \cite{HeathWearable:15} where the main lobe gain $G_N$ is $N$ with beam width $\theta_N=2\pi/\sqrt{N}$, while neglecting side lobes. The transmitted mmW signal experiences path loss attenuation with the exponent $\alphaM >2$ as well as Rayleigh fading with unity mean. Note that LOS dependent mmW scenarios may be more in line with Rician fading, which incorporates the number of LOS paths, or Nakagami-$m$ fading that can be a mean value approximation of the Rician fading by adjusting its shape factor $m$. The Rayleigh model in this paper, nevertheless, enables better tractability while providing at least a lower bound of the maximized DL rate under Nakagami-$m$ fading. That is because the signal-to-interference-plus-noise ($\SINR$) with Nakagami-$m$ fading is lower bounded by the value with Rayleigh fading, verified via a stochastic ordering technique \cite{JLee:14}.

At a typical DL user and at an UL BS, the mmW $\SINR$ is then given as:

\vspace{-10pt}\small\begin{align}
\SINR_{m.d} &:=  \frac{ G_N P_{m.d}  g r^{-\alpha_\text{m}} \mathds{1}_{\RL}(r) }{\sum_{i\in \PhiActM } G_N \Theta_i P_{m.d}g_i {r_i}^{-\alpha_\text{m}} \mathds{1}_{\RL}(r_i)+ \sigma^2}\\
\SINR_{m.u} &:=  \frac{G_N P_{m.u} g r^{-\alpha_\text{m}} \mathds{1}_{\RL}(r) }{\sum_{i\in \PhiActU } G_N \Theta_i P_{m.u}g_i {r_i}^{-\alpha_\text{m}} \mathds{1}_{\RL}(r_i)+ \sigma^2}
\end{align} \normalsize
\noindent where $r$ and $r_i$ respectively denote the distances to the associated BS (or user) and interfering BSs (or users) from the origin, $\sigma^2$ noise power, and $\Theta_i$ the probability that the $i$-th transmitter with beam width $\theta$ interferes with the typical receiver. Note that $\Theta_i$ is be assumed to be a uniformly random variable since the associated receiver location of the $i$-th interfering transmitter is also uniformly distributed.

Now that our interest of this paper is confined to ultra-dense cellular networks, we hereafter only consider signal-to-interference-ratio ($\SIR$) instead of $\SINR$ by neglecting $\sigma^2$ under a interference-limited regime, unless otherwise noted.

\begin{figure*}
\centering
 	\subfigure[$\mu$W SE ($\alphaMu = 4$)]{\includegraphics[width=9cm]{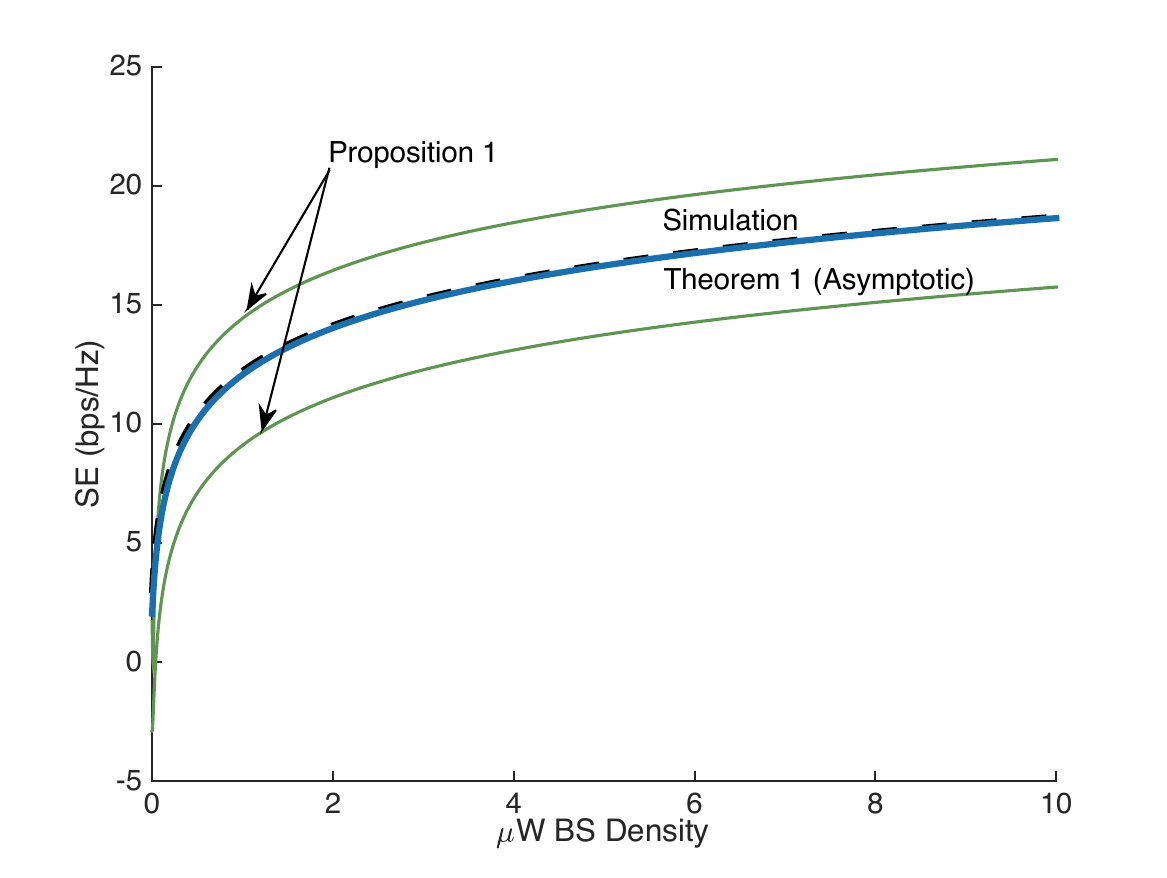}} \label{Fig:muWBounds} \hspace{-.5cm}
	\subfigure[mmW SE ($\theta = 15^\circ$, $\RL = 10$ m, $\alphaM = 2.5$)]{\includegraphics[width=9cm]{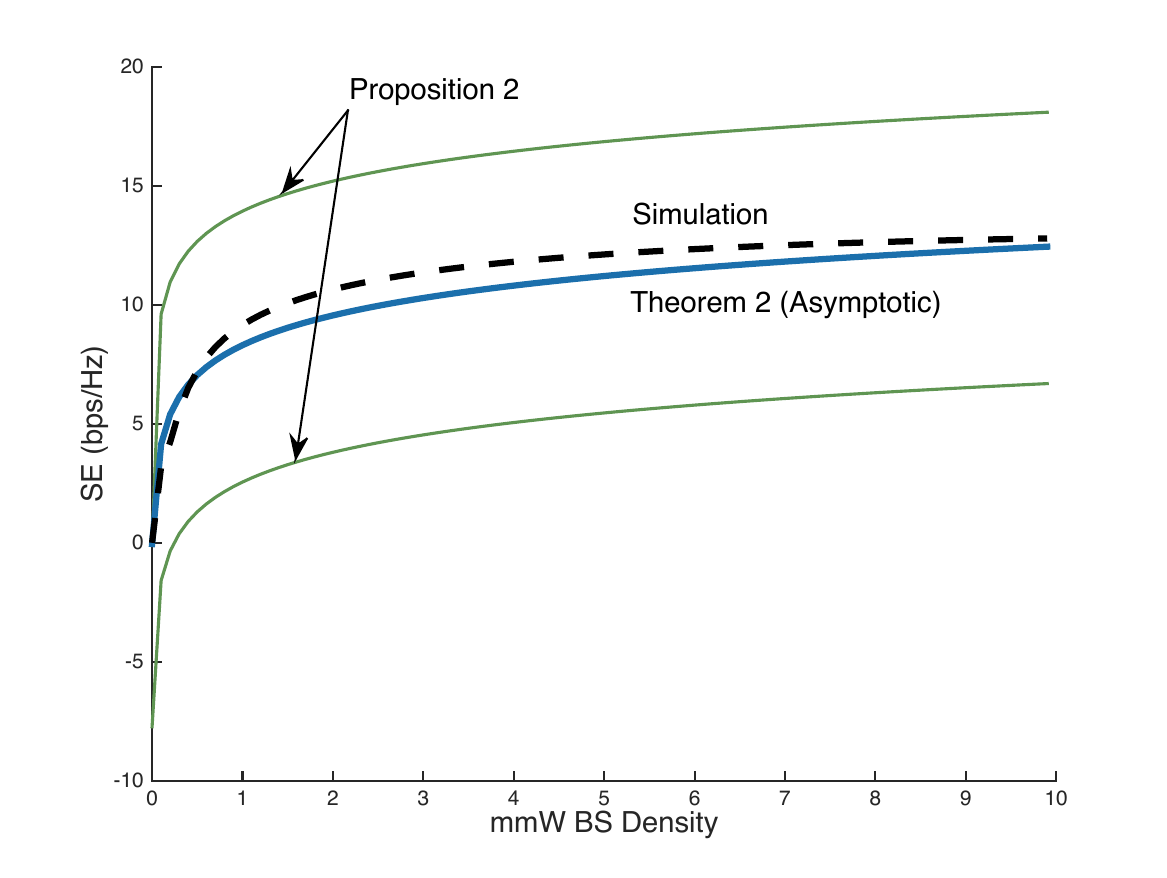} } \label{Fig:mmWBounds}
	\caption{The mm-$\mu$W DL/UL UDN SE upper and lower bounds as well as asymptotic values when $\lambdaU = 0.01$.} \label{Fig:SEBounds}
\end{figure*}

\subsection{Resource Management under Limited mmW UL Bandwidth} \label{Sect:Sys_Resource}
Consider multiple access of mm-$\mu$W users by default follows orthogonal frequency-division multiple access (OFDMA) during each unit time slot, unless otherwise noted. The entire mm-$\mu$W spectrum bandwidths are given respectively as $W_m$ and $W_{\mu}$.

For mmW, DL/UL resources are allocated in a time-division duplex (TDD) manner via adjusting the mmW UL allocation ratio $\beta_m$. While mmW DL operates with the entire $W_m$, mmW UL operations only utilize $\WMUL < \WM$ due to the PAPR outage at mobile users in return for the use of huge $W_m$ (see Fig.~\ref{Fig:Network}).

Under using single-carrier frequency-division multiple access (SC-FDMA) at mmW UL transmissions, the said UL PAPR bottleneck can be relaxed. Nevertheless, $\WMUL$ is still delimited due to huge sampling rate at mmW transmissions. According to Nyquist-Shannon theorem, the sampling rate at DAC should satisfy at least $2 \WMUL$. The bottleneck then becomes the power limitation at DAC whose dissipated power increases along with the sampling rate, or $\WMUL$ \cite{SamsungDAC:14}. Specific mmW UL PAPR outage constraint and DAC power limitation are deferred to Section \ref{Sect:LimitedUL}.

For $\mu$W, DL/UL resource allocation follows TDD via adjusting $\mu$W UL allocation ratio $\beta_\mu$. This model is also able to capture FDD based $\mu$W operations whose UL spectrum allocation is $\beta_\mu W_\mu$. In any cases, $\mu$W bandwidth $\WMu$ is much smaller than $\WM$, and therefore $\mu$W UL operations are assumed to be free from the PAPR outage and DAC power limitation at mobile users.

\section{Spectral Efficiencies of Millimeter-Wave Overlaid Ultra-Dense Cellular Networks}  \label{Sect:SE}
The closed-form representations for mm-$\mu$W SEs are of prime concerns in this section. The derived results will play a salient role in tractable mm-$\mu$W resource management in Section \ref{Sect:RscMngmt}. The SE is defined as ergodic capacity $\E \log[1 + \SINR]$, in units of nats/sec/Hz (1bit $\approx$ 0.693 nats) unless otherwise noted. Unlike $\mu$W SE, mmW SE incorporates the degradation resulting from blockages. Its impact is captured via average LOS distance $\RL$. We hereafter let the subscripts $m$ and $\mu$ denote mmW and $\mu$W respectively. The path loss exponent $\alpha$ and BS density $\lambda$ without subscripts can be either mmW or $\mu$W, which improves notational reusability. In the following subsections, closed-form mm-$\mu$W SEs are first presented in Section \ref{Sect:SEresult}, and then their derivations are described in the remainder of the subsections.

\subsection{Closed-form mm-$\mu$W UDN SEs} \label{Sect:SEresult}

For a BS density $\lambda$, let $\hat{\lambda}$ denote $\lambda / \lambdaU$. This ratio of BS-to-user density determines the ultra-densification of a network as follows.

\begin{definition}\emph{(UDN) }
A cellular network with BS density $\lambda$ is called a UDN where $\hat{\lambda} \gg 1$. The notation $f\gg g$ implies f is sufficiently large so that $O\(g/f\)$ is approximated by $0$.
\end{definition}

In a UDN regime, the difference between DL and UL $\mu$W SEs, denoted respectively by $\gammaMuDL$ and $\gammaMuUL$, is negligible. That is because most of the associated user-BS distances of both DL and UL become identical. Such a UDN regime, furthermore, leads to the asymptotic convergence of the upper and lower bounds of $\gammaMuDL$ (or $\gammaMuUL$) as $\lambdahMu$ increases. Its derivation is deferred to Proposition 1 in Section \ref{Sect:SEmuW}. The corresponding asymptotic $\mu$W UDN SE is presented in the theorem below.

\setcounter{theorem}{0}
\begin{theorem}\emph{(Asymptotic $\mu$W UDN SE)  }
For $\lambdahMu \rightarrow \infty$, DL and UL $\mu$W SEs identically converge to $\gammaMu$, i.e.
\emph{$\gammaMu = \lim\limits_{\lambdahMu \rightarrow \infty}\gammaMuUL = \lim\limits_{\hat{\lambda}_\mu \rightarrow \infty}\gammaMuDL$}, and the value is given as:
\begin{eqnarray} 
&&\hspace{-20pt} \gammaMu = \frac{\alphaMu}{2} \log \lambdahMu. \hfill 
\end{eqnarray}
\end{theorem}

In a similar manner, mmW DL and UL SEs are denoted respectively by $\gammaMDL$ and $\gammaMUL$, and their values become identical in a UDN regime. The upper and lower bounds of $\gammaMDL$ (or $\gammaMUL$) converge as specified at  Proposition 3 in Section \ref{Sect:SEmmW}. The converged mmW UDN SE is stated in the following theorem.

\setcounter{theorem}{1}
\begin{theorem}\emph{(Asymptotic mmW UDN SE)} For $\lambdahM \rightarrow \infty$, DL and UL mmW SEs identically converge to $\gammaM$, i.e.
$\gammaM = \lim\limits_{\lambdahM \rightarrow \infty}\gammaMUL = \lim\limits_{\lambdahM \rightarrow \infty}\gammaMDL$, and the value is given as:
\emph{\begin{align}
\gammaM &= {{\frac{\alphaM \pL}{2}}}   \log\lambdahM 
\end{align}}
where \emph{$\pL := 1 - \exp\(-\lambdaM \pi \RL^2 \)$}.
\end{theorem}

The results are illustrated in Fig.~\ref{Fig:SEBounds}. According to Theorems~1 and 2, both mm-$\mu$W SEs have two remarkable characteristics under a UDN regime as follows.

\begin{remark}\emph{(BS-to-User Density Ratio Dependence)} UDN SE logarithmically increases with BS-to-user density ratio.
\end{remark}
This first finding emphasizes what matters in a UDN is the ratio of BS density to user density, also highlighted via simulation in \cite{TowardUDN:15}. In addition, it is worth noticing that the result above is distinct from the well-known SE behavior that is independent of BS densification under an interference-limited regime \cite{Andrews:2011bg}. Such a difference is caused by unique characteristics of UDN interference that is limited by user density. In a DL UDN, for instance, 
user density puts a cap on interference increase since it determines whether a BS becomes an interferer via turning on/off the BS. On the other hand, BS densification increases the desired signal power via reducing the association distance by $1/(2\sqrt{\lambda})$ on average, resulting in the logarithmic SE improvement.

The second finding includes the exposition of a newfound \emph{weak-interference regime} that has a mixed nature of interference/noise-limited behaviors as follows.
\begin{remark}\emph{(Weak Interference-Limited UDN)} An interference-limited UDN shows distinctive behavior toward mmW blockages and BS densification, which is similar to the trends in a noise-limited regime.
\end{remark}

\begin{figure*}
\setcounter{equation}{11}
\begin{align}
\log\( 1 + \[ {\rho_{\mu}}^{-1} \lambdahMu \]^{{\frac{\alphaMu}{2}}}\)  -\frac{\alphaMu}{2}  \leq \; \gammaMuDL \text{ (or $\gammaMuUL$)} \; \leq  \log\( 1 + \[ \( 1 + \frac{2}{\alphaMu}\)\lambdahMu\]^{\frac{\alphaMu}{2}}\) - \frac{\alphaMu}{2} \label{Eq:Prop1}
\end{align}
\hrulefill
\end{figure*}

Clear understanding of this unique UDN trait follows from revisiting an interference-limited network whose noise power is negligible compared to interference, i.e. $\SINR \approx \SIR$. In a traditioanl interference-limited network, it is well known that SE increases along with path loss exponent \cite{Andrews:2011bg} and blockages \cite{Heath:13}. Their corresponding interference reductions dominate the desired power decreases, thereby leading to SE increases. Under this interference-limited network, in addition, SE is independent of BS densification and transmission power increase \cite{Andrews:2011bg}. The reason is they improve the desired signal power and interference alike, thus cancelling their impact each other.

In an interference-limited UDN, the effects of path loss exponent and transmission power increase directly follow the same trends in a traditional interference-limited network. On the other hand, the increase in blockages, i.e. increasing $\pL$ via reducing $\RL$, decreases mmW UDN SE. That is because a large number of mmW UDN interferers dominate the blocked interferers. It makes the interference reduction due to blockages negligible. The UDN SE therefore depends solely on the desired signal power decrease due to blockages. Unlike a traditional interference-limited network, it is also remarkable that BS densification increases UDN SE in a logarithmic manner as stated after Remark 1. Such trends toward blockages and BS densification under an interference-limited UDN are analogous to the behaviors in a traditional noise-limited network where interference is negligible compared to noise.

When reminding of the UDN interference delimited by user density, it is worth mentioning that a UDN with low user density may become noise-limited. Nevertheless, its trivial tendency due to removing interfering interaction is not of our interest, thus neglected. The following subsections describe the major techniques for the derivations of Theorems 1 and 2. The relationship of the techniques are summarized in Fig. 3.


\subsection{Preliminary Techniques} \label{Sect:SEPre}

This section introduces two approximations that are feasible under a UDN regime. These techniques make a detour around the obstacles disrupting the tractable SE analysis. 

The first technique tackles the nonhomogeneity of the interferers, active BSs for DL and active users for UL. As described in Section \ref{Sect:Network}, this intractability is induced by the random active user selection at multiple access. The selection probability of active users depends on each BS's cell coverage whose size is affected by its neighboring BSs; for instance, a user in a BS dense region is more likely to be selected due to the small cell coverage size of the associated BS. This inter-node dependency results in non-homogeneous interferer distributions.

In a UDN regime, however, such a location dependency diminishes since cell size tends to be identical. We thus approximate the interferers are independently thinned. The thinning probability for DL is active BS probability $p_a$, and for UL is user selection probability $p_s$, derived in \cite{Yu2011} as follows.
\setcounter{equation}{6}

\vspace{-5pt}\small\begin{align}
p_a &= 1-\[1 + \(3.5\hat{\lambda}{}\)^{-1}\]^{-3.5} \\
p_s &= p_a \hat{\lambda}
\end{align}\normalsize

\begin{figure}
\centering
	\includegraphics[width=9cm]{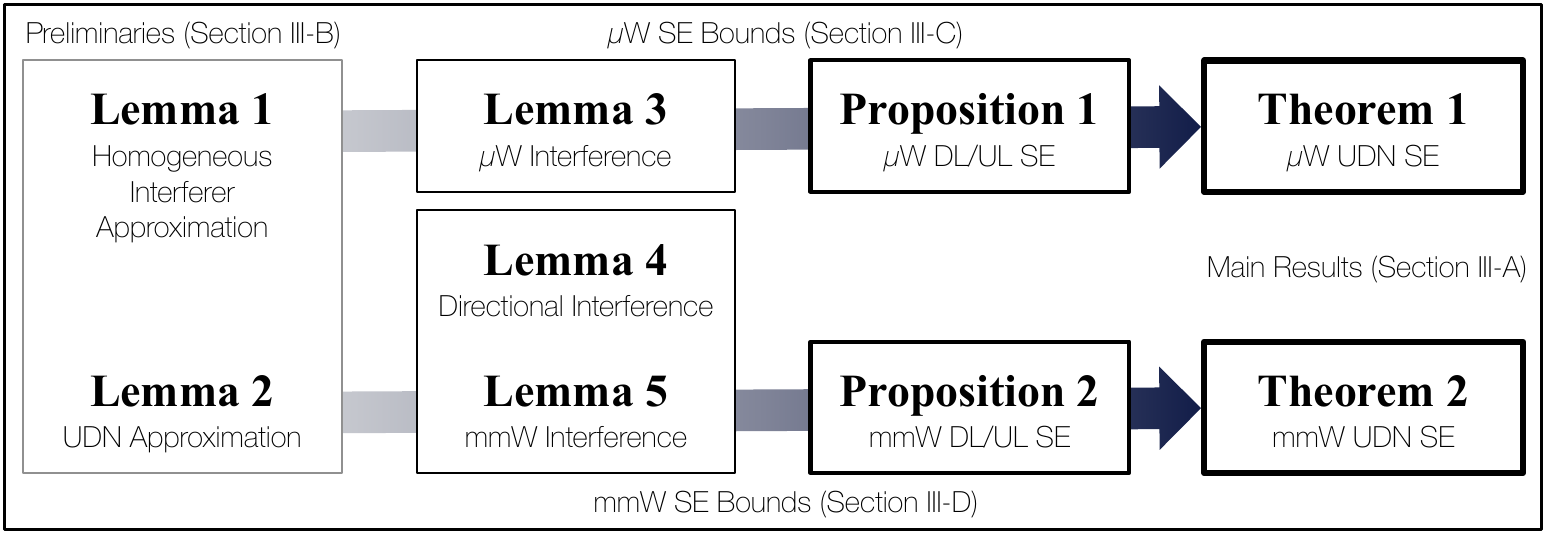}
	\caption{Mathematical technique connections toward the derivations of Theorems 1 and 2.} \label{Fig:Techniques}
\end{figure}

The corresponding interferer distribution is specified in the following lemma.
\begin{lemma}\emph{(Homogeneous Interferer Approximation)} As $\hat{\lambda}\rightarrow~\infty$, both DL and UL interferer coordinates identically converge to a homogeneous PPP with density $\lambdaU$.
\end{lemma}
This underpins the homogeneous PPP approximation of the interferer coordinates in a UDN, of which the validity is also investigated by simulation in \cite{SLeeKHuang12}. The result intuitively implies that the interferer coordinates converge to user coordinates in a UDN regime. It is thus notable that such a homogeneous PPP approximation is valid only when users are uniformly distributed. Investigating the impact of different user distributions in a UDN could therefore be an important issue for further directions.

The second technique enhances the tractability of the SE representations. As an example of the SE calculation, consider a DL UDN with BS density $\lambda$. At a typical user, the associated BS is the nearest BS out of the entire BSs deployed with density $\lambda$ while the interferers are active BSs whose density is approximated as $\lambdaU$ (or $p_a \lambda$) via Lemma 1. Exploiting Campbell theorem \cite{StoyanBook:StochasticGeometry:1995} with the same technique at the equation (16) in \cite{Andrews:2011bg}, the corresponding DL UDN SE $\gamma_{\text{d}}$ is represented as

\vspace{-5pt}\small\begin{align}
\gamma_{\text{d}} & = \int_{t>0} \E_R \[  \exp\( - \lambdaU \pi  (e^t - 1)^{\frac{2}{\alpha} } R^2 \int_{(e^t-1)^{-\frac{2}{\alpha}}}^\infty  \frac{du}{1 + u^{\frac{\alpha}{2}}}   \) \] dt \label{Eq:PfProp1ExactPre}
\end{align}\normalsize
where $R$ denotes the distance to the associated BS from the typical user. 

Nevertheless, this representation is not tractable due to the innermost integration range that depends on $t$. To remove such a dependency, we first exploit the following approximation proposed in our preliminary work \cite{JHPark:14}, which is tight in a UDN regime.

\begin{lemma}\emph{(UDN Approximation)} Consider a UDN with BS density $\lambda$. For a nonnegative constant $I\ll \lambda$, the following approximation holds.
\begin{align}
\E_{R}\[ \exp\( -\lambdaU \pi r^2 I \)\] &\approx \[ 1 - \hat{\lambda}{}^{-1} I  \]^+
\end{align}
\end{lemma} 
In the following Sections \ref{Sect:SEmuW} and D, this result will be combined with \eqref{Eq:PfProp1ExactPre}, and leads to tractable upper and lower bounds of $\mu$W and mmW DL/UL SEs, each of which identically converges on the asymptotic closed-form SE respectively in Theorems 1 and 2.

\begin{figure*}
\setcounter{equation}{15}
\small\begin{align} 
\int_{t>0} \pLt \( 1 - \rhoM \lambdahM^{-1} \[ N^{-\frac{1}{2}}(e^t-1)\]^{\frac{2}{\alphaM}}\)^+ dt   \leq &\;\; \gammaMDL \text{ (or $\gammaMUL$)} \;\;
\leq \int_{t>0} \pLt \( 1- \[\(1 + \frac{2}{\alphaM}\) \lambdahM\]^{-1} \[N^{-\frac{1}{2}}\(e^t - 1\) \]^{\frac{2}{\alphaM}} \)^+ dt  \label{Eq:Prop2}\\ 
\pL \[ \log\( 1 + N^{\frac{1}{2}}\[ {\rhoM}^{-1}\lambdahM \]^{{\frac{\alpha}{2}}}\)  -\frac{\alphaM}{2} \] \leq &\;\; \gammaMDL \text{ (or $\gammaMUL$)} \;\;
 \leq C_{L_2} \log\( 1 + N^{\frac{1}{2}}\[ \(1+\frac{2}{\alphaM}\)\lambdahM \]^{{\frac{\alphaM}{2}}}\) \label{Eq:Corollary2} \\
&\hspace{-7cm}\text{where $\pLt:=  1 -  e^{ -\lambdaM\pi \RL^2 \lb 1 + \lambdahM^{-1}\rhoM \[ N^{-\frac{1}{2}}(e^t-1)\]^{\frac{2}{\alphaM}}   \rb }$ and $C_{L_2} := 1 - e^{-\lambdaM \pi \RL^2\lb1 + \rhoM \( 1 + \frac{2}{\alphaM}\) \rb}$} \nn
\end{align}\normalsize
\hrulefill
\end{figure*}

\subsection{$\mu$W SE Bounds (Derivation of Theorem 1)} \label{Sect:SEmuW}

Let $\rhoMut$ denote $\mu$W interference constant defined as $\int_{(e^t-1)^{-\frac{2}{\alphaMu}}}^\infty  \frac{du}{1 + u^{\frac{\alphaMu}{2}}}$. This corresponds with the innermost integration in \eqref{Eq:PfProp1ExactPre}. To avoid double integration in \eqref{Eq:PfProp1ExactPre} for tractability, we consider the lower and upper bound of $\rhoMut$ as below.

\setcounter{equation}{10}
\begin{lemma}\emph{($\mu$W Interference Constant Bounds)} The following inequality holds for all $t>0$
\begin{align}
\(1 + \frac{2}{\alphaMu} \)^{-1} - (e^t - 1)^{-\frac{2}{\alphaMu}}  \; \leq \; \rhoMut \; \leq \; \rhoMu
\end{align}
where $\rho_{\mu} :=  \int_{u>0} \frac{du}{1 + u^{\frac{\alphaMu}{2}} } = \(\frac{2 \pi}{\alphaMu}\)\text{csc}\(\frac{2 \pi}{\alphaMu}\)$.
\end{lemma}

Applying Lemmas 2 and 3 to \eqref{Eq:PfProp1ExactPre} provides the following tractable upper and lower bounds of $\mu$W DL/UL UDN SEs.

\begin{proposition}\emph{($\mu$W DL/UL UDN SE Bounds)}
In a UDN regime, $\mu$W DL SE \emph{$\gammaMuDL$} and UL SE are identically bounded as \eqref{Eq:Prop1} at the top of the page.
\end{proposition}\setcounter{equation}{12}
This result reveals that $\mu$W DL/UL SEs share the same upper and lower bounds in a UDN regime. The reason is because the user-to-interfering active BS distances asymptotically converge on the inter-user distances as BS density increases.

In addition, the proposition shows the upper and lower bounds of $\mu$W DL/UL UDN SEs logarithmically increases with the BS-to-user density ratio. For the DL, the improvement in SE results from the fact that BS densification reduces its corresponding Voronoi cell sizes. This turns off the BSs having no active user within their cells. Consequently, at a typical user, the interfering active BSs decrease, and their density ends up with being delimited by user density (recall Lemma 1). At the same time, the densification shortens the distance between a typical user and its associated BS, thereby yielding the SE increase.

For the UL, the SE improvement follows from the cell size reduction along with densification in a similar way to the DL case. The difference is the BS densification may increase interferers by increasing the number of the selected users. In a UDN regime, nevertheless, such an interferer increasing effect rarely occurs, leading to improve the SE as in the DL scenario.

\subsection{mmW SE Bounds (Derivation of Theorem 2)} \label{Sect:SEmmW}

Directional mmW signal transmissions decrease the aggregate interference at a typical receiver compared to omnidirectional transmissions. The reduced interference amount is specified as below.

\setcounter{equation}{12}
\begin{lemma}\emph{(Directional Interference Thinning)}
At a typical user (or BS), DL mmW aggregate interference $I_{\Sigma_{\text{m.d}}}$ and UL mmW aggregate interference $I_{\Sigma_{\text{m.u}}}$ are represented as:
\begin{align}
I_{\Sigma_{\text{m.d}}} &\approx \sum_{i \in \Phi_{\text{m}}\( \lambdaU N^{-\frac{1}{\alphaM}}\)} g_i {r_i}^{-\alphaM} \mathds{1}_{\RL}(r_i) \\
I_{\Sigma_{\text{m.u}}} &\approx \sum_{i \in \Phi_{\text{u}}\( \lambdaU N^{-\frac{1}{\alphaM}}\)} g_i {r_i}^{-\alphaM} \mathds{1}_{\RL}(r_i).
\end{align}
\end{lemma}

The rest of the derivation follows the similar procedure for the $\mu$W case in Section \ref{Sect:SEmuW}. Let $\rhoMt := \int_{ (e^t-1)^{-\frac{2}{\alphaM}}}^{\(\frac{\RL}{r}\)^2 (e^t-1)^{-\frac{2}{\alphaM}}}  \frac{du}{1 + u^{\frac{\alphaM}{2}}}$. The following lemma provides tractable upper and lower bounds of $\rhoMt $.

\begin{lemma}\emph{(mmW Interference Constant Bounds)} The following inequality holds: (i) for $\lambdahM \rightarrow \infty$ almost surely; and (ii) for a UDN with probability $1-\exp\(-\rho_m \lambda_u \pi \RL^2\)$.
\begin{align}
1 - \(1 + \frac{\alphaM}{2}\)^{-1}-\(e^t-1\)^{-\frac{2}{\alphaM}}  \; \leq \; \rhoMt  \; \leq \; \rhoM
\end{align}
where $\rhoM: = \int_{u>0} \frac{du}{1 + u^{\frac{\alphaM}{2}}} = \(\frac{2 \pi}{\alphaM}\)\text{csc}\(\frac{2 \pi}{\alphaM}\)$
\end{lemma}
In a UDN, it is highly feasible to apply the inequality above for LOS distance and/or user density. According to Table I, our target geography corresponds with such an environment. At Gangnam, for instance, the inequality holds with probability greater than $0.93$ ($\lambda_u = 1 \times 10^{-4}$, $\alpha_m = 2.5$). We thus assume Lemma 5 always holds in a UDN hereafter.

The mmW DL/UL UDN SEs are then upper and lower bounded by applying Lemmas 4 and 5 to \eqref{Eq:PfProp1ExactPre}.
\begin{proposition}\emph{(mmW DL/UL UDN SE Bounds)}
In a UDN regime, mmW DL SE \emph{$\gammaMDL$} and UL SE $\gammaMUL$ are identically bounded as \eqref{Eq:Prop2} at the top of this page.
\end{proposition} \setcounter{equation}{16}
It shows both DL/UL mmW SEs have the same upper and lower bounds as in the case of $\mu$W SE in Proposition 1. The impact of ultra-densification however cannot be explicitly interpreted in this form due to the integrations therein. 

To capture the mmW UDN behavior in a more intuitive way, we consider less tight but more tractable upper and lower bounds. The modified lower bound follows from $\pL \leq \pLt$. For the upper bound, observe that the maximum $t$ is $\log\( 1 + N^{\frac{1}{2}}\[\(1 + \frac{2}{\alphaM}\)\lambdahM \]^{\frac{\alphaM}{2}} \)$ such that the upper bound integrand is non-zero. Combining them provides the following closed-form upper and lower bounds.

\begin{corollary} \emph{(Tractable mmW DL/UL UDN SE Upper Bound)} In a UDN regime, mmW DL SE $\gammaMDL$ and UL SE $\gammaMUL$ are identically upper bounded as \eqref{Eq:Corollary2} at the top.
\end{corollary}\setcounter{equation}{17}
This indicates mmW SE also logarithmically increases with the BS-to-user density ratio as in the $\mu$W SE in Section \ref{Sect:SEmuW}. The only difference is its average LOS probability $\pL$ due to the blockage vulnerability of mmW signals. 

\section{Uplink/Downlink Resource Management in Millimeter-Wave Uplink Decoupled Ultra-Dense Cellular Networks}	\label{Sect:RscMngmt}
This section jointly optimizes mm-$\mu$W DL/UL resource allocations so that the overall average DL rate is maximized while guaranteeing a minimum UL rate. In addition, we propose a mmW UL decoupling that further enhances the DL rate without procuring additional spectrum resources. For the calculations including mm-$\mu$W SEs, we henceforth approximate the exact values by the results in Theorems 1 and 2, of which the tightness is validated in Fig.~\ref{Fig:SEBounds}. The following subsections firstly specify preliminaries and problem formulation, and then provide the mm-$\mu$W resource allocation results.

\subsection{Limited mmW UL Bandwidth} \label{Sect:LimitedUL}

The mm-$\mu$W resource management incorporates the limited mmW UL spectrum amount $\WMUL$ as discussed in Section \ref{Sect:Sys_Resource}. Such a problem follows from (i) PAPR outage constraint and (ii) DAC power limitation at mobile users.

Firstly, the PAPR outage constraint is given as $\Pr\( \PAPR> \delta \) \leq \epsilon$ for constants $\delta, \epsilon>0$. According to \cite{Jiang:08}, the $\PAPR$ outage constraint can be rephrased as

\vspace{-5pt}\small\begin{align}
\Pr\( \PAPR> \delta \) &= 1 - \exp\( - \frac{w e^{-\delta}}{f_s} \sqrt{\frac{\pi \delta}{3}} \)\leq \epsilon \label{Eq:PAPR}
\end{align}\normalsize
where $w$ denotes UL spectrum bandwidth and $f_s$ subcarrier spacing.

Secondly, the limitation on DAC power dissipation $\Pdac$ is given as $\Pdac \leq P_{\max}$. As in the case of analog-to-digital converter (ADC) \cite{SundeepADCmmW:15}, $\Pdac$ is assumed to be linearly proportional to the use of spectrum bandwidth. This yields the $\Pdac$ limitation as

\vspace{-5pt}\small\begin{align}
\Pdac &= \cdac w \leq P_{\max} \label{Eq:Pdac}
\end{align}\normalsize
where $\cdac$ is a non-negative constant in units of W/Hz.

These enable calculating the limited mmW UL bandwidth $\WMUL$, the maximum $w$ simultaneously satisfying \eqref{Eq:PAPR} and \eqref{Eq:Pdac}. The smaller value from either PAPR outage constraint or DAC power limitation determines $\WMUL$ as follows.

\vspace{-5pt}\small\begin{align}
W_{m.u} &= \min\l\{ f_s e^{\delta}\sqrt{\frac{3}{\pi \delta}}\log\(1-\epsilon\)^{-1}, P_{\text{max}}/\cdac, W_m \r\} \label{Eq:mmWLimitedUL} 
\end{align}\normalsize

Note that PAPR outage is likely to occur under OFDMA based mmW UL transmissions. SC-FDMA may relax this outage problem, but its excessive DAC dissipation may still put a strain on the use of the entire mmW UL bandwidth. The limited mmW UL bandwidth in \eqref{Eq:mmWLimitedUL} capture both scenarios, and its impact is investigated by simulation in Section \ref{Sect:NumericalAlloc}.

\subsection{mmW UL Decoupling}  \label{Sect:mmWULDecouple}
The limited mmW UL bandwidth makes it difficult to satisfy the minimum UL to DL rate ratio, and thereby holds back overall DL rate increase. To resolve this problem, we propose  a novel \emph{mmW UL decoupling} scheme that enables mmW UL receptions at the incumbent $\mu$W BSs (see user 2 in Fig. 1). By the aid of this proposed approach, mmW UL users are able to associate with not only mmW but also $\mu$W BSs; in other words, mmW UL decoupled BS to user density ratio becomes $\lambdahM + \lambdahMu$ while mmW DL BS to user density ratio is $\lambdahM$. This as a result increases mmW spatial reuse as well as mmW received signal power, leading to ease the UL bottleneck. 

Note that the DL and UL in the proposed scheme are decoupled over different mm-$\mu$W RATs. Especially in an UDN, its association distance reduction via obtaining more UL BSs is the sole advantage. It is thus different from $\mu$W single RAT UL/DL decoupling \cite{Decouple:14,BoccardiDUde:15} that resolves the DL/UL imbalance in transmission power and/or scheduling, of which the both mismatches are negligible under UDNs (see Definition 1 and discussion after Remark 2).

From a practical point of view, such a mmW UL decoupling is cost-effective. The approach at legacy $\mu$W BSs only requires additional mmW receive antennas and corresponding circuit components without power amplifiers. Sharing the existing data backhauls in part while having no need of additional power suppliers and site acquisition cost are expected to make the solution practically viable. The impact of mmW UL decoupling on mm-$\mu$W resource management and its resultant DL rate improvement are provided in Section \ref{Sect:Opt_WandDens3}.

\subsection{SE in Sparse mmW Networks}
To further investigate the initial mmW deployment stages, this subsection additionally considers mmW SE under a sparse network. For simplicity, a sparse mmW network where $\lambdahM < 1$ is assumed to be noise-limited. The desired mmW SE is then given as follows.
\begin{proposition}\emph{(mmW Sparse SE)} For $\lambdahM \leq 1$, DL mmW SE $\gammaMDL$ and UL SE $\gammaMUL$ are:
\small\begin{align}
\gammaMDL & = \nn\\
&\hspace{-5pt}2\pi \lambdahM^2 \lambda_u{P_{m.d}}^{\frac{2}{\alphaM}} \int_0^{\RL} r e^{\frac{\sigma^2r^{\alphaM}}{P_{m.d}N}-\lambdahM\lambda_u \pi r^2}\Gamma\(0,\frac{\sigma^2 r^{\alphaM}}{P_{m.d}N}\)dr \label{Eq:mmWSEsparseDL} \\ 
\gammaMUL & = \nn\\
&\hspace{-5pt}2\pi \lambdahM^2 \lambda_u {P_{m.u}}^{\frac{2}{\alphaM}}\int_0^{\RL} r e^{\frac{\sigma^2r^{\alphaM}}{P_{m.u}N}-\lambdahM\lambda_u \pi r^2}\Gamma\(0,\frac{\sigma^2 r^{\alphaM}}{P_{m.u}N}\)dr \label{Eq:mmWSEsparseUL}
\end{align}\normalsize
where $\Gamma(0,x):= \int_x^\infty e^{-t}/t dt$. 
\end{proposition}
Note that the results above incorporate the impact of multiple user access. Under a uniformly random scheduler, the average selection probability of a typical user by a mmW BS is $\lambdahM$, multiplied in front of the integrations in \eqref{Eq:mmWSEsparseDL} and \eqref{Eq:mmWSEsparseUL}. 

Let $\gammaMDL^{(s)}(\lambdahM)$ and $\gammaMUL^{(s)}(\lambdahM)$ respectively denote \eqref{Eq:mmWSEsparseDL} and \eqref{Eq:mmWSEsparseUL} under sparse mmW networks. We hereafter consider $\gammaMDL^{(s)}$ and $\gammaMUL^{(s)}$  as the DL and UL mmW SEs for $\lambdahM < 1$ and the results in Theorems 1 and 2 for $\lambdahM\geq 1$.


\subsection{Problem Formulation} \label{Sect:RscProblem}
Define DL and UL average rates $\RDL$ and $\RUL$ as follows.

\vspace{-5pt}\small\begin{eqnarray}
	\RDL &:=& \( 1-\beta_m \)W_m \gammaMDL + \( 1-\beta_\mu\) \WMu \gammaMu  \label{Eq:RateDL} \\
	\RUL &:=& \beta_m \WMUL \gammaMUL + \beta_\mu \WMu \gammaMu  \label{Eq:RateUL}
\end{eqnarray}\normalsize

Consider following two assumptions: \textsf{A1}. mmW DL (or UL) capacity with the entire spectrum use exceeds the $\mu$W DL (or UL) capacity, i.e. $W_m \gamma_m > W_\mu \gamma_\mu$; and \textsf{A2}. the required mimimum UL rate is no greater than DL rate, i.e. $\zeta \leq 1$ where $\zeta$ denotes the minimum ratio of UL to DL rates. The former \textsf{A1} is justified by the motivation of the mmW use, enabling higher capacity via huge $W_m$ that compensates low $\gamma_m$ due to high distance attenuations. The later \textsf{A2} is based on the well-known traffic statistics that DL demand is much higher that UL demand. 

Such \eqref{Eq:RateDL} and \eqref{Eq:RateUL} are then calculated under a given $\mu$W UDN and the overlaid mmW sparse/ultra-dense network therein. Note that $\mu$W may also be a sparse network in practice although its ultra-densification is expected to be less demanding compared to a mmW UDN from both technical and economic perspectives. Analyzing the impact of $\mu$W is directly viable by utilizing the SE representation under a sparse $\mu$W network in our preceding result, equation (4) in \cite{JHPark:14}. In this paper, however, a sparse $\mu$W network is not addressed in order to focus more on the impact of mmW BS densification.

We hereafter focus on mm-$\mu$W UL resource allocations via adjusting the UL allocation ratios $\beta_m$ and $\beta_\mu$ that are no greater than unity. DL mm-$\mu$W allocation ratios are straightforwardly determined as $1-\beta_m$ and $1-\beta_\mu$. The objective problem is then formulated as follows.
\begin{align}
      \textsf{P1. }&
   \begin{aligned}[t]
    & \underset{ \beta_m,\; \beta_\mu }{\text{maximize}} \; \RDL  \notag\\
   \end{aligned}   \notag\\
   &\text{subject to} \notag \\
   & \quad   \RUL/\RDL   \geq  \zeta \label{Eq:ConstULQoS} \\
   &\quad 0 \leq \beta_m,\beta_\mu \leq 1. \label{Eq:ConstULMax}
\end{align} 

The problem above is a linear programming with two variables under a feasible region given by \eqref{Eq:ConstULQoS} and \eqref{Eq:ConstULMax}. The optimal solutions of the problem and their engineering implications are elucidated in the following section.

\subsection{mm-$\mu$W Resource Management} \label{Sect:Opt_WandDens2}

Our resource management aims at maximizing the overall mm-$\mu$W DL rate with guaranteeing the minimum UL/DL rate ratio via adjusting mm-$\mu$W UL allocations. The corresponding solution is provided as mmW BS-to-user density ratio $\lambdahM$ increases. 

The optimal mm-$\mu$W resource allocations for \textsf{P1} are given as below.
\begin{proposition} \emph{(Optimal mm-$\mu$W UL Resource Allocations)} In a $\mu$W UDN, optimal mm-$\mu$W UL allocations $\(\beta_m^*, \beta_\mu^*\)$ are given as follows.
\begin{itemize}
\item For $\lambdahM < 1$,
\small\begin{align}
\hspace{-25pt}\begin{array}{ll}
\( 0, \[1 + \zeta^{-1}\]^{-1}\[1 + \frac{2 \WM \gammaMDL^{(s)}(\lambdahM)}{\alphaMu \WMu \log \lambdahMu}  \] \)  &\hspace{-15pt}\text{if $\lambdahM < \lambdahS$} \vspace{3pt}\\
\( \[\zeta + \frac{\WMUL \gammaMUL^{(s)}(\lambdahM)}{\WM \gammaMDL^{(s)}(\lambdahM)} \]^{-1} \[ \zeta - \frac{\alphaMu\WMu \log \lambdahMu}{2\WM \gammaMDL^{(s)}(\lambdahM)}\] ,1 \)  &\hspace{-5pt}\text{otherwise} \label{Eq:Proposition3}
\end{array}
\end{align}\normalsize
\item For $\lambdahM \geq 1$,
\small\begin{align}
\hspace{-25pt}\begin{array}{ll}
\( 0, \[1 + \zeta^{-1}\]^{-1}\[1 + \frac{\alphaM \WM \pL \log \lambdahM}{\alphaMu \WMu \log \lambdahMu} \] \)  &\text{if $\lambdahM < \lambdahU$} \vspace{3pt}\\
\( \[\zeta + \frac{\WMUL}{\WM} \]^{-1} \[\zeta - \frac{\alphaMu \WMu \log \lambdahMu}{\alphaM \WM \pL \log \lambdahM} \] ,1 \)  &\quad\text{otherwise} \label{Eq:Proposition3}
\end{array}
\end{align}\normalsize
\end{itemize}
where $\lambdahU = \lambdahMu^{\frac{\alphaMu \WMu}{\zeta \alphaM \pL \WM}}$ and $\lambdahS$ is the value satisfying $\gammaMDL^{(s)}(\lambdahM)=\WMu \gammaMu /\(\zeta \WM\)$
\end{proposition} 
It is notable that the mm-$\mu$W UL allocations make a sudden change when crossing a critical mmW BS to user density ratio $\lambdahS$ or $\lambdahU$ for sparse or ultra-dense mmW networks. The following remark specifies this behavior.
 
\begin{remark} \emph{($\mu$W-to-mmW UL Allocation)} As mmW BS density increases,
\begin{enumerate}[ \;1.]
\item $\mu$W resource should first be allocated to UL while keeping mmW dedicated to DL ($\beta_m^* = 0$);
\item mmW UL allocation begins to increase after $\mu$W UL allocation reaches the entire spectrum ($\beta_\mu^* = 1$).
\end{enumerate}
\end{remark}
Note that such a sequential $\mu$W-to-mmW resource allocation always occurs so long as the UL mmW bandwidth is limited, i.e. $\WMUL < \WM$.

The remark implies allocating $\mu$W resource more to UL than to DL for sufficiently dense mmW BS deployment. This ends up with the DL dedicated mmW and UL dedicated $\mu$W scenarios under extensive mmW BS proliferation, which is in the opposite direction to the traditional $\mu$W resource allocation trend giving top priority to DL.

Such DL mmW and UL $\mu$W dedications are interestingly in accordance with the recently proposed mm-$\mu$W resource allocation concepts from industry and regulatory perspectives \cite{SamsungmmWave:11, BoccardiDUde:15}. Providing stable UL communications for control signals \cite{SamsungmmWave:11} and mitigating mmW UL transmission power limitation due to mmW's high electromagnetic field exposure \cite{BoccardiDUde:15} are their reasons behind. From a different point of view, our proposed approach results from detouring mmW UL bandwidth limitation.

Next, the tractable result in Proposition 4 provides the minimum $\mu$W UL allocation at the initial stage of mmW deployment.
\begin{corollary} \emph{(Min. $\mu$W UL Allocation)} The minimum $\mu$W UL allocation amount depends solely on $\zeta$, i.e. $\lim_{\lambdahM \rightarrow 0} \beta_\mu^* = 1/\(1 + \zeta^{-1}\)$.
\end{corollary}
The results shows the minimum $\beta_\mu^*$ ranges from $0$ to $0.5$ for $\zeta$'s from $0$ to $1$. This straightforwardly accords with the resource allocation tendency in a traditional cellular network as the impact of mmW overlay is neglected.

Similarly, Proposition 4 provides the maximum mmW UL allocation at the extremely dense mmW BS deployment.
\begin{corollary} \emph{(Max. mmW UL Allocation)} The maximum mmW UL allocation amount depends only on $\zeta$, $\WM$, and $\WMUL$, i.e. $\lim_{\lambdahM \rightarrow \infty} \beta_m^* = \( 1 + \frac{\WMUL}{\zeta \WM}\)^{-1}$.
\end{corollary}
It is remarkable that the maximum $\beta_m^*$ does not depend on $\mu$W bandwidth and SE but on mmW DL and UL bandwidths. This is because the mmW UL rate becomes dominant under this extreme mmW BS densification.

Lastly, Proposition 4 tractably captures the blockage impact on mm-$\mu$W resource allocations via average LOS probability $\pL$, an increasing function of average LOS distance $\RL$.

\begin{remark} \emph{(Blockage Impacts on Resource Allocations)} In a mm-$\mu$W UDN,
\begin{enumerate}[1.]
\item Both mm-$\mu$W UL allocations decrease as blockage becomes more severe;
\item The $\mu$W UL allocation gaps between geographical locations increase for $\lambdahM < \lambdahU$ whereas the gaps decrease for $\lambdahM \geq \lambdahU$.
\end{enumerate}
\end{remark}
The first argument results from the weak interference-limited characteristics on blockages in a UDN (see Remark 2). More blockages decrease mmW UDN SE. This results in larger DL rate decrement than UL rate due to $\WMUL<\WM$. As a consequence, DL/UL rate asymmetry becomes less severe.

The second argument follows from the $\mu$W-to-mmW UL allocation order (Remark 3). When $\lambdahM < \lambdahU$, the mmW SE improvement by densification amplifies the blockage dependent impact on the SE, leading to the result. When $\lambdahM \geq \lambdahU$, on the other hand, mmW BS densification becomes almost always achieving LOS conditions, i.e. $\pL\rightarrow 1$, and therefore the blockage impact on SE rapidly diminishes. Fig.~\ref{Fig:RscwoDC} in Section \ref{Sect:NumericalAlloc} illustrates such a blockage dependent behavior.

The optimal mm-$\mu$W UL allocations in Proposition 3 are followed by the maximized DL rate with the minimum UL constraint \eqref{Eq:ConstULQoS} in \textsf{P1} as below.

\begin{proposition} \emph{(Max. DL UDN Rate)} For a mm-$\mu$W UDN with the optimal UL allocation given in Proposition 3, the maximized DL rate $\RDL^*$ is:
\begin{align}
\begin{array}{ll}
\[2\(\zeta + 1 \)\]^{-1} \log \lambdahMu^{\alphaMu \WMu}\lambdahM^{\alphaM \WM \pL}   & \text{if $\lambdahM < \lambdahU$}; \vspace{3pt}\\
\[2\(\zeta + \frac{\WMUL}{\WM} \)\]^{-1} \log \lambdahMu^{\alphaMu \WMu} \lambdahM^{\alphaM \WMUL \pL}  & \text{otherwise}. 
\end{array}
\end{align}
\end{proposition}
The result shows that the maximized DL rate is a logarithmic function of mm-$\mu$W BS-to-user densities. It is remarkable that the maxmized DL rate becomes highly restricted by the limited UL bandwidth $\WMUL$ for $\lambdahM \geq \lambdahU$. This can be relieved via mmW UL decoupling proposed in Section \ref{Sect:mmWULDecouple}. The following subsection investigates its impact on mm-$\mu$W resource allocations.

\subsection{mm-$\mu$W Resource Management with mmW UL Decoupling} \label{Sect:Opt_WandDens3}
Under mmW UL decoupled systems, legacy $\mu$W BSs can receive mmW signals. This makes mmW UL BS density to user ratio become $\lambdahM + \lambdahMu$. This leads to the modified mm-$\mu$W UL resource allocations.
\begin{proposition} \emph{(Optimal mm-$\mu$W UL Resource Allocations with mmW UL Decoupling)} In a $\mu$W UDN, optimal mm-$\mu$W UL allocations $\(\beta_m^*, \beta_\mu^*\)$ are given as follows.
\begin{itemize}
\item For $\lambdahM < 1$:
\begin{align}
\hspace{-25pt}\small\begin{array}{ll}
\( 0, \[1 + \zeta^{-1}\]^{-1}\[ 1 + \frac{2 \WM \gammaMDL^{(s)}(\lambdahM)}{\alphaMu \WMu \log \lambdahMu} \] \)  &\hspace{-20pt} \text{if $\lambdahM < \lambdahS$}; \vspace{3pt}\\
\( \[\zeta + \frac{\WMUL \gammaMUL^{(s)}(\lambdahM+\lambdahMu)}{\WM\gammaMDL^{(s)}(\lambdahM)} \]^{-1} \[ \zeta - \frac{\alphaMu \WMu \log\lambdahMu}{2 \WM \gammaMDL^{(s)}(\lambdahM)}\] ,1 \)  &\hspace{-10pt}\text{otherwise} \label{Eq:Proposition3}
\end{array}\normalsize
\end{align}\vspace{-10pt}
\item For $\lambdahM \geq 1$:
\begin{align}
\hspace{-25pt}\small\begin{array}{ll}
\( 0, \[1 + \zeta^{-1}\]^{-1}\[1 + \frac{\alphaM \WM \pL \log \lambdahM}{\alphaMu \WMu \log \lambdahMu} \] \)  &\hspace{-45pt}\text{if $\lambdahM < \lambdahU$}; \vspace{3pt}\\
\( \[\zeta + \frac{\WMUL \log\(\lambdahM + \lambdahMu\)}{\WM \log \lambdahM} \]^{-1}\[\zeta - \frac{\alphaMu \WMu \log \lambdahMu}{\alphaM \WM \pL \log \(\lambdahM + \lambdahMu\)} \] ,1 \)  \\ 
\hspace{200pt}\text{otherwise} \label{Eq:Proposition3} 
\end{array}\normalsize
\end{align}
\end{itemize}
\end{proposition} 
Note that the results above assume $W_m \gammaMDL > \WMUL \gammaMUL$ always holds. Its opposite case indicates the DL dedicated mmW rate is less than the mmW UL rate with the limited UL mmW bandwidth. This can mathematically occur but not in practice, thus neglected. 

The resource allocation impact of mmW UL decoupling compared to the allocations without the decoupling is specified in the following remark.
\begin{remark} \emph{(mmW UL Decoupling Impact on Resource Allocations)} mmW UL decoupling reduces mmW UL allocation when $\lambdahM$ exceeds $\lambdahS$ or $\lambdahU$ respectively for a sparse or ultra-dense mmW network.
\end{remark}
The UL allocation of mmW bandwidth only occurs when $\lambdahM$ exceeds $\lambdahS$ or $\lambdahU$. For less $\lambdahM$, mmW bandwidth is dedicated to DL transmissions, yielding such a result.

In addition, the DL rate resulting from the mm-$\mu$W allocation in Proposition 6 is presented as below.
\begin{proposition} \emph{(Max. DL UDN Rate with mmW UL Decoupling)} For a mm-$\mu$W UDN with mmW UL decoupling and the optimal UL allocation given in Proposition 5, maximized DL rate $\RDL^*$ is:
{\begin{align}\small
\hspace{-8pt}\begin{array}{ll}
\[2\(1 + \zeta\)\]^{-1}\log \lambdahMu^{\alphaMu \WMu}\lambdahM^{\alphaM \WM \pL}    \hfill \text{if $\lambdahM < \lambdahU$}; \vspace{5pt}\\
\[2\(\zeta + \frac{\WMUL \log\( \lambdahM + \lambdahMu\)}{\WM \log\lambdahM} \)\]^{-1} \log\lambdahMu^{\alphaMu \WMu}  \( \lambdahM + \lambdahMu\) ^{\alphaM \WMUL \pL}   \\
\hfill \text{otherwise}.
\end{array} 
\end{align}\normalsize }
\end{proposition}
This DL rate with mmW decoupling and its improvement from the DL rate without the decoupling are visualized at Fig.~\ref{Fig:Rate} in Section \ref{Sect:NumericalAlloc}.

\begin{comment}
\begin{figure*}
\centering \hspace{-1cm}
 	\subfigure[Gangnam (2 $\times$ 2 km$^{2}$)]{\includegraphics[width=6cm]{Figs_M/Map_Gangnam.pdf}   }  \hspace{-.2cm}
 	\subfigure[Jongro (1 $\times$ 1 km$^{2}$)]{\includegraphics[width=5.8cm]{Figs_M/Map_Jongro.pdf}    }\hspace{-.2cm}
	\subfigure[Yonsei (2 $\times$ 2 km$^{2}$)]{\includegraphics[width=5.6cm]{Figs_M/Map_Yonsei.pdf}    }
	\caption{Target areas for building statistics: Gangnam, Jongro, and Yonsei, Seoul, Korea \cite{JHParkGeo:15}.}
\end{figure*}
\end{comment}

\begin{table*}[!t]
\renewcommand{\arraystretch}{1.3}
\caption{Average LOS Distances Measured from Building Statistics in Seoul, Korea}
\label{Table:GeoStat}
\centering
\begin{tabular}{l | r | r | r | r | r}
\hline
 &\textbf{Gangnam} & \textbf{Jongro}& \textbf{Yonsei} &Manhattan \cite{Kulkarni:14}&Chicago \cite{Kulkarni:14}\\
Building parameters (unit) &$\(2\times2\text{ km}^2\)$&$\(1\times1\text{ km}^2\)$&$\(2\times2\text{ km}^2\)$&$\(1\times1\text{ km}^2\)$&$\(0.5\times0.5\text{ km}^2\)$\\
\hline\hline
Avgerage perimeter $\rho$ (m$^2$) &59.02&39.29&51.99&73.78&114.48\\
Avgerage area $A$ (m$^2$) &218.60&107.67&173.95&312.26&886.46\\
Coverage $\kappa$ (\%) &34.77&46.90&25.48&45.83&42.02\\
Avgerage height $\E H$ (m) &14.23 & 8.12 & 11.14 & 101.00 & 28.95 \\
Height $\sim$ \text{log$\mathcal{N}$}$(\mu$,$\sigma)$ & (1.62, 0.27) & (0.69, 0.55)& (1.10, 0.34)& (3.32, 0.30) & (1.36, 1.23)\\
\hline
2D blockage parameter $\beta$ & 0.073 & 0.014 & 0.056 & 0.092 & 0.045 \\
3D height parameter $\eta$ & 0.36 & 0.22 & 0.13 & 0.12 & 0.46  \\
\textbf{2D LOS distance $R_L^{2D}$} (m) & \bf17.77 & \bf7.22 & \bf26.63 & \bf11.75 & \bf25.88  \\
\textbf{3D LOS distance $\RL$} (m) & \bf{49.61} & \bf{33.33} & \bf198.76 & \bf98.11 & \bf56.20  \\
\hline
\end{tabular}
\end{table*}

\section{Numerical Evaluation with a Real Geography Based  3D Blockage Model} \label{Sect:Numerical}
The mm-$\mu$W SEs in Section \ref{Sect:SE} and the corresponding DL/UL resource management in Section \ref{Sect:RscMngmt} are numerically evaluated in this section. To maximize their practical feasibility, we apply a 3D blockage model, and calculate average LOS distance $\RL$ with the actual building geography in Seoul, Korea.

\subsection{Average LOS Distance under a Real Geography Based 3D Blockage Model} \label{Sect:3DBlockage}
The possibility that a given link with distance $r$ is LOS not only depends on blockage locations but also their heights. We consider both effects by applying a 3D blockage model \cite{Heath:13}. Its corresponding 3D average LOS distance is given as:

\small\begin{equation}
\RL = \frac{2(1-\kappa)}{\beta \eta }
\end{equation}\normalsize
where 
\vspace{-5pt}\small\begin{eqnarray}
\beta &:=& \frac{- 2 \rho \log\( 1 - \kappa\)}{\pi A},\\
\eta &:=& \int_{0}^{1} \Pr\( H \leq (1-s) B\)ds, \label{Eq:BlockageHeight}
\end{eqnarray}\normalsize
$\rho$ average blockage perimeter, $\kappa$ average building coverage, $A$~average building area, $H$ building height, and $B$ BS height. Note that putting $\eta = 1$ into \eqref{Eq:BlockageHeight} yields 2D average LOS distance $R_L^{\text{2D}}$.

By the aid of the Ministry of Land, Infrastructure, and Transport of Korea, we calculate $\beta$ and $\eta$ that correspond to three hotsopt regions in Seoul, Korea: Gangnam, Jongro, and Yonsei university town, visuallized in Fig. 4 (see \cite{JHParkGeo:15} for their geographical data and further calculation details). The results are presented in Table \ref{Table:GeoStat}. Comparing 2D and 3D LOS distances reveals that these two values are less correlated. For instance, the $R_L^{2D}$ in Gangnam is larger than the value in Manhattan, but such a relationship is reversed from the perspective of $\RL$. it is thus necessary to consider a 3D blockage model for a practically feasible analysis.

\begin{figure}
\centering
 	\vspace{-15pt}\includegraphics[width=9cm]{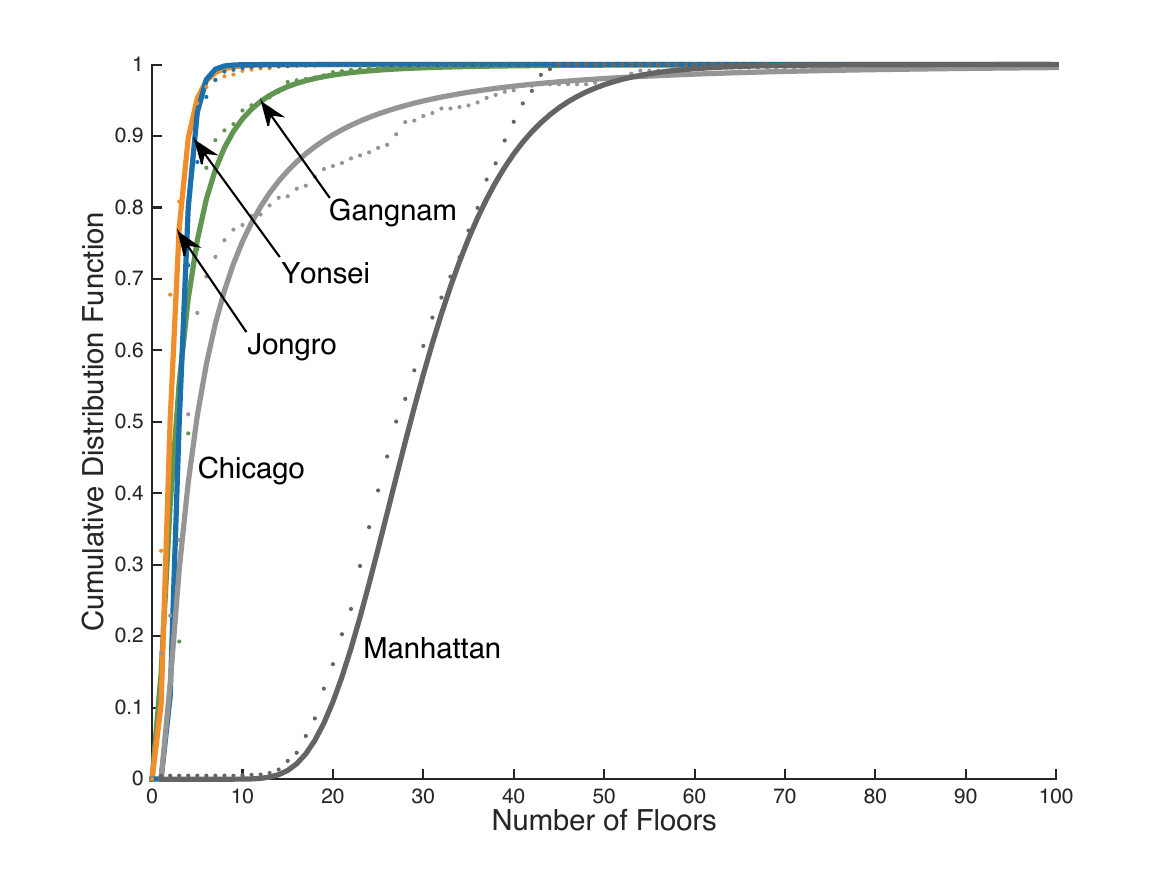}
	\caption{Distributions of the number of floors in Seoul, Korea (dotted lines), and their lognormal fitted curves (solid lines). The root mean square error of each fitted curve is less than $0.016$.} \label{Fig:FLDistrib}
\end{figure}

\begin{figure}
\centering
	\vspace{-15pt}\includegraphics[width=9cm]{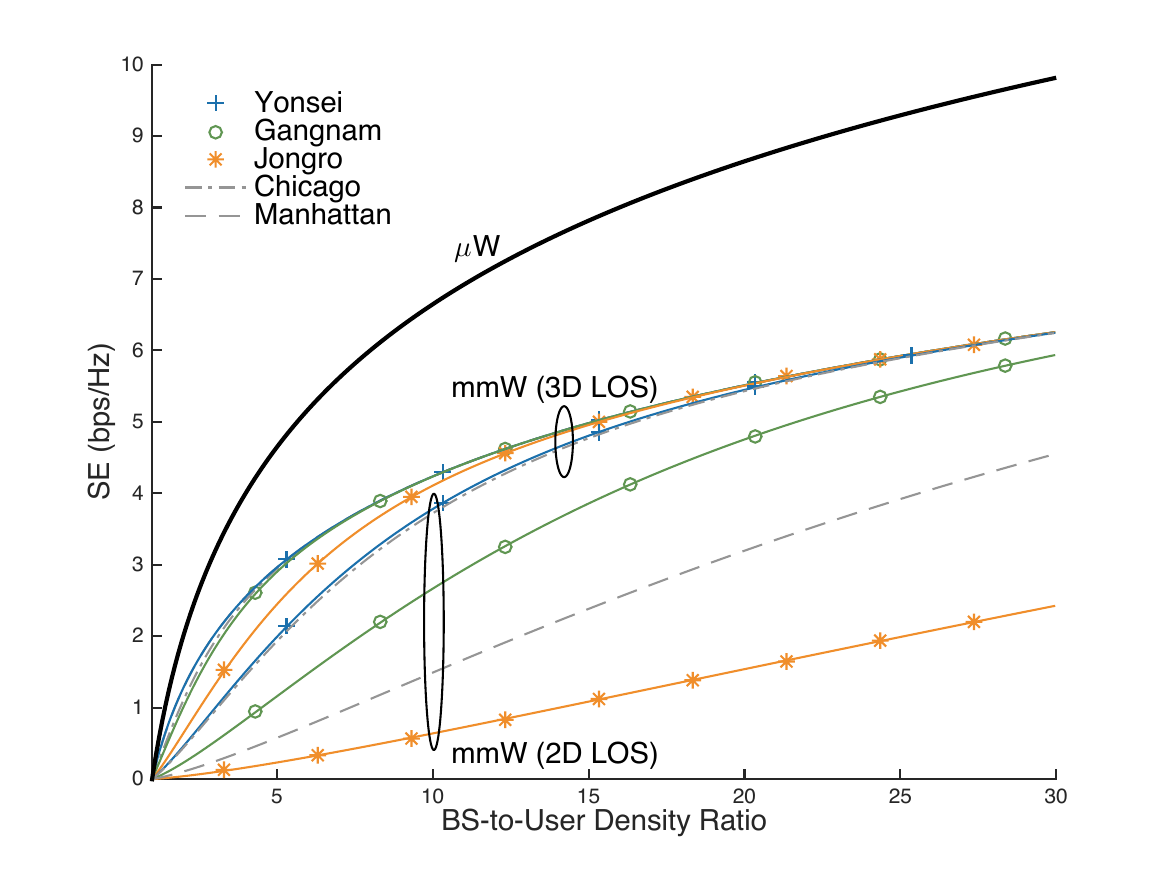}
	\caption{The spectral efficiencies of DL $\mu$W and DL mmW with the 2D/3D LOS distance values at Gangnam, Jongro, Yonsei in Seoul, Korea as well as Manhattan and Chicago in the US where the LOS distance values are found in Table I ($\alphaMu = 4$, $\alphaM = 2.5$, $\lambdaU = 1$).} \label{Fig:SEMUMM}
\end{figure}

The calculations of $\beta$ rely on QGIS, an open source geographic information system (GIS) \cite{QGIS}. The process for $\eta$ additionally requires the distribution of building height $H$ due to \eqref{Eq:BlockageHeight}. Now that the given geographic data does not include the building height information, we detour this problem via the information on the number of building floors. We assume unit floor height is $3$ m, and then derive the building floor distribution via the data curve fitting (see Fig.~\ref{Fig:FLDistrib}). For BS heights, also required in \eqref{Eq:BlockageHeight}, we assume each BS height follows the average building height $\E H$. Note that the $\eta$ values of Manhattan and Chicago are also calculated in this work for fair comparisons of the blockage environments. Their geographical data follow from the same sources in \cite{Kulkarni:14}.

\subsection{mm-$\mu$W SEs under a Real Geography Based 3D Blockages} \label{Sect:NumericalSE}

Based on a 3D blockage model with the actual building statistics summarized in Table I, Fig.~\ref{Fig:SEMUMM} visualizes the mm-$\mu$W SEs in Theorems 1 and 2. The significant gap between mm-$\mu$W UDN SEs can be explained via the effects of path loss exponent and mmW blockages in a weak interference-limited regime. The smaller mmW path loss exponent for LOS links results in the lower mmW SE than $\mu$W SE. Further mmW SE degradation results from the more blockages than the $\mu$W environment that is blockage-free as assumed in Section \ref{Sect:SysBlockage}.

Another notable point is the impact of 3D blockage model whose resultant mmW SEs are higher than the values under a 2D blockage model. At a receiver in the 2D model, all buildings whose locations are in the straight line from the transmitter become blockages. In the 3D model, on the other hand, different building heights make some of these buildings still guarantee LOS conditions. Such effects reduces blockage degradations compared to the 2D model, yielding higher SEs.

\subsection{mm-$\mu$W Resource Management}\label{Sect:NumericalAlloc}

We by default consider mm-$\mu$W bandwidths are respectively given as $2$ GHz and $20$ MHz under $25\%$ minimum UL/DL rate ratio. The corresponding result is compared with the case where mmW bandwidth is set as $3$ GHz under the $50\%$ UL/DL rate ratio. Note that such UL/DL rate ratios are in line with the 3GPP Release 12 standards \cite{Rel12:2015}: for its peak rate specifications, the UL/DL rate ratio $25\%$ indicates user equipment category (CAT) 5, 10, and 14; the ratio $50\%$ is in accordance with CAT 1--3, and 8, where the ratio for smart devices ranges from $8\%$ to $50\%$.

The remaining parameters are given as follows: $\lambdaU = 10^{-4}$, $\lambdaMu = 2 \lambdaU$, $P_{\mu.d}=P_{m.d}=40$ dBm, $P_{\mu.u}=P_{m.u}=20$ dBm; (for PAPR) $f_s = 244.14$ kHz \cite{WRoh:14}, $\delta = 12$ dB, $\epsilon = 0.01$; (for DAC power) $c_w=24.28$ mW/GHz \cite{AkiraDAC:13}, $P_{\text{max}}=20$ mW \cite{SundeepADCmmW:15}.

\begin{figure}
\vspace{0pt}\includegraphics[width=9cm]{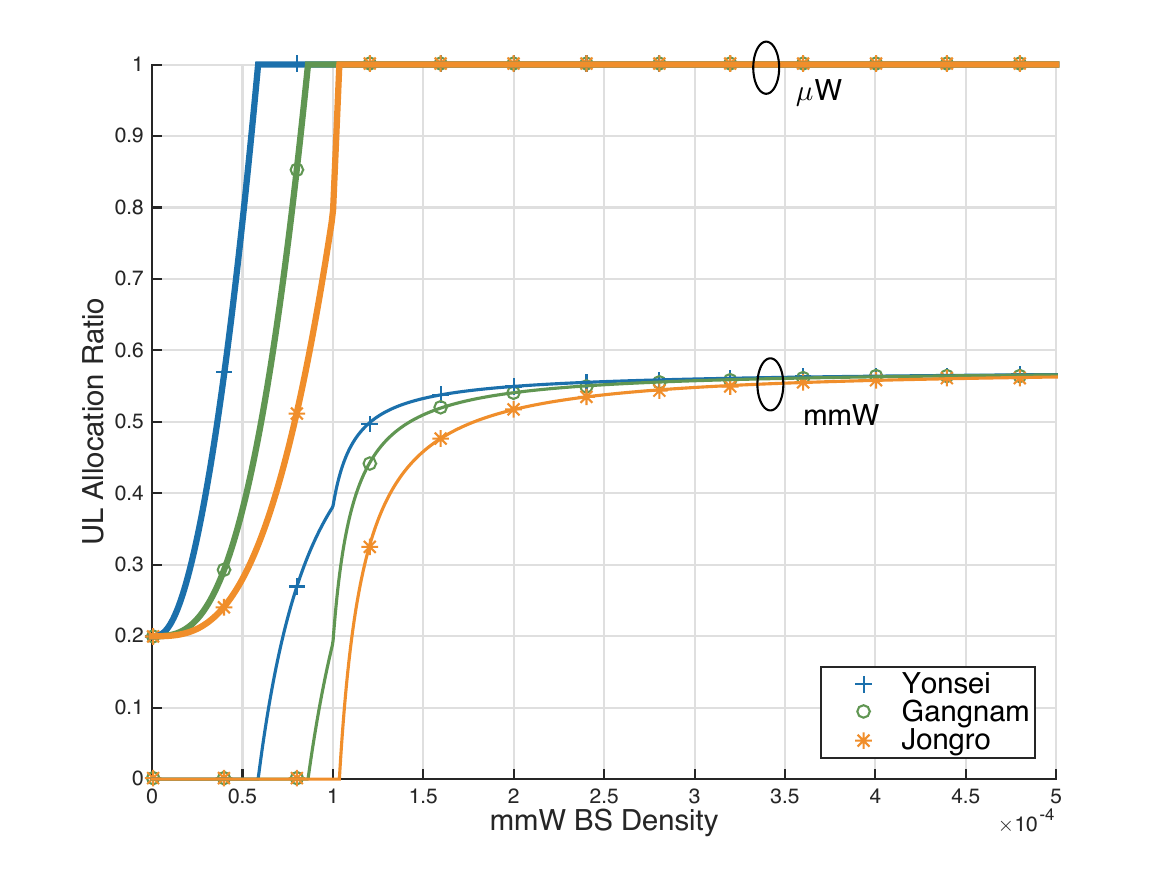}  
\caption{Optimal mm-$\mu$W UL resource allocations ($W_m = 2$ GHz, $\zeta = 0.25$) }   \label{Fig:RscwoDC}
\end{figure}

Fig.~\ref{Fig:RscwoDC} visuallizes mm-$\mu$W UL allocations without mmW UL decoupling (see Proposition 4 for analytic details). As mmW BS density increases, it shows $\mu$W UL allocation increases while keeping mmW UL allocation zero, i.e. DL dedicated mmW. This trend holds until from $0.57$ (Yonsei) to $1.03$ (Jongro) times mmW BS densification compared to user density, where it starts dedicating $\mu$W to UL. After reaching the UL dedication of $\mu$W, mmW UL allocation then begins to increase (Remark 3). For mmW allocations, abrupt curve slope transitions at $\lambda_m = 1 \times 10^{-4}$ (or $\lambdahM = 1$) occur, where mmW BS densification changes from sparse to ultra-dense regimes. For both mm-$\mu$W allocations, high blockage locations ($\RL$: Yonsei$>$Gangnam$>$Jongro) require more UL allocations since their DL/UL rate mismatch becomes severe (Remark 4).

\begin{figure}
\vspace{0pt}\includegraphics[width=9cm]{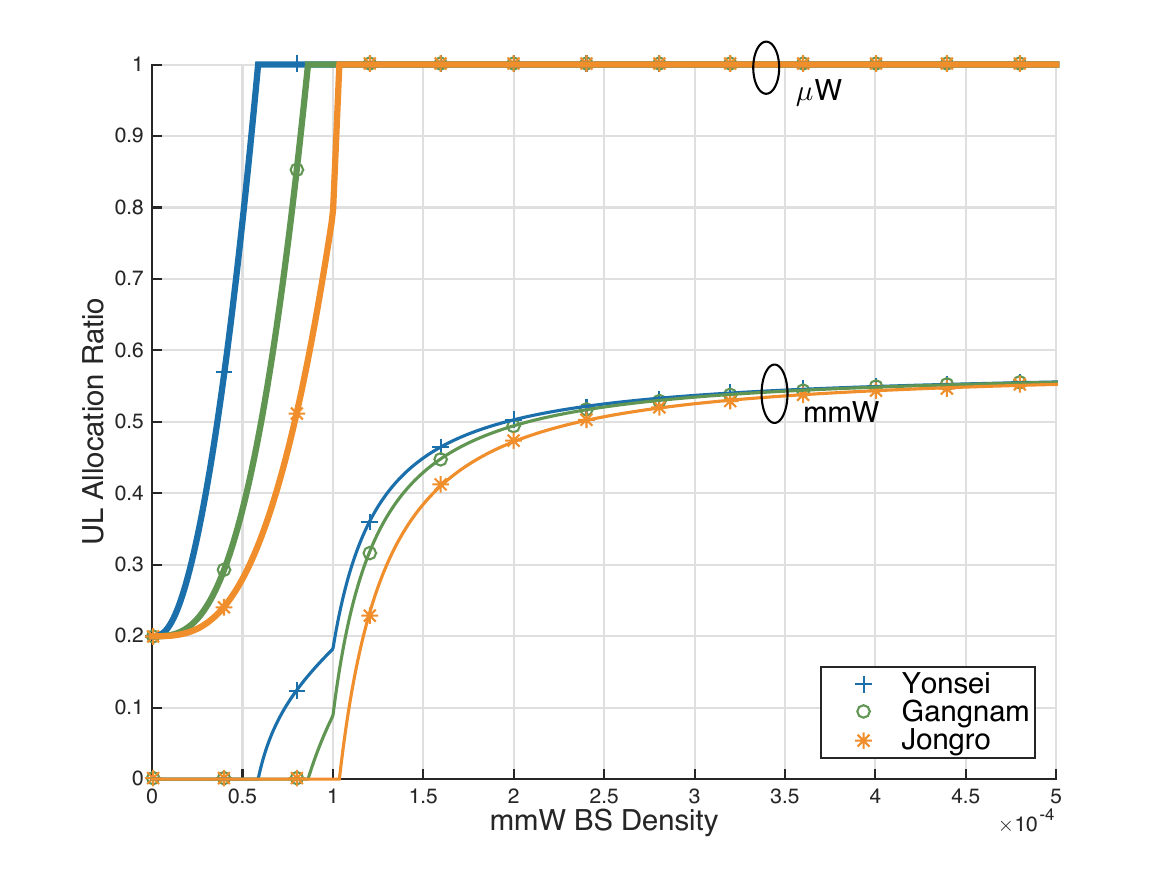}  
\caption{ Optimal mm-$\mu$W UL resource allocations with mmW UL decoupling ($W_m = 2$ GHz, $\zeta = 0.25$)  } \label{Fig:RscwDC}
\end{figure}

Fig.~\ref{Fig:RscwDC} illustrates the impact of mmW UL decoupling on mm-$\mu$W UL resource allocations (Proposition 6). The proposed UL decoupling reduces mmW UL allocations (Remark 5). For Yonsei, as an example, $47$\% mmW UL allocation reduction is achieved at $\lambdahM = 1$. For all locations, it is worth noting that mmW UL decoupling yields the identical maximum mmW UL allocations $0.55$ to the case without the decoupling since the value is independent of mmW BS densification as well as blockages (Corollary 3).

\begin{figure}
\vspace{0pt}\includegraphics[width=9cm]{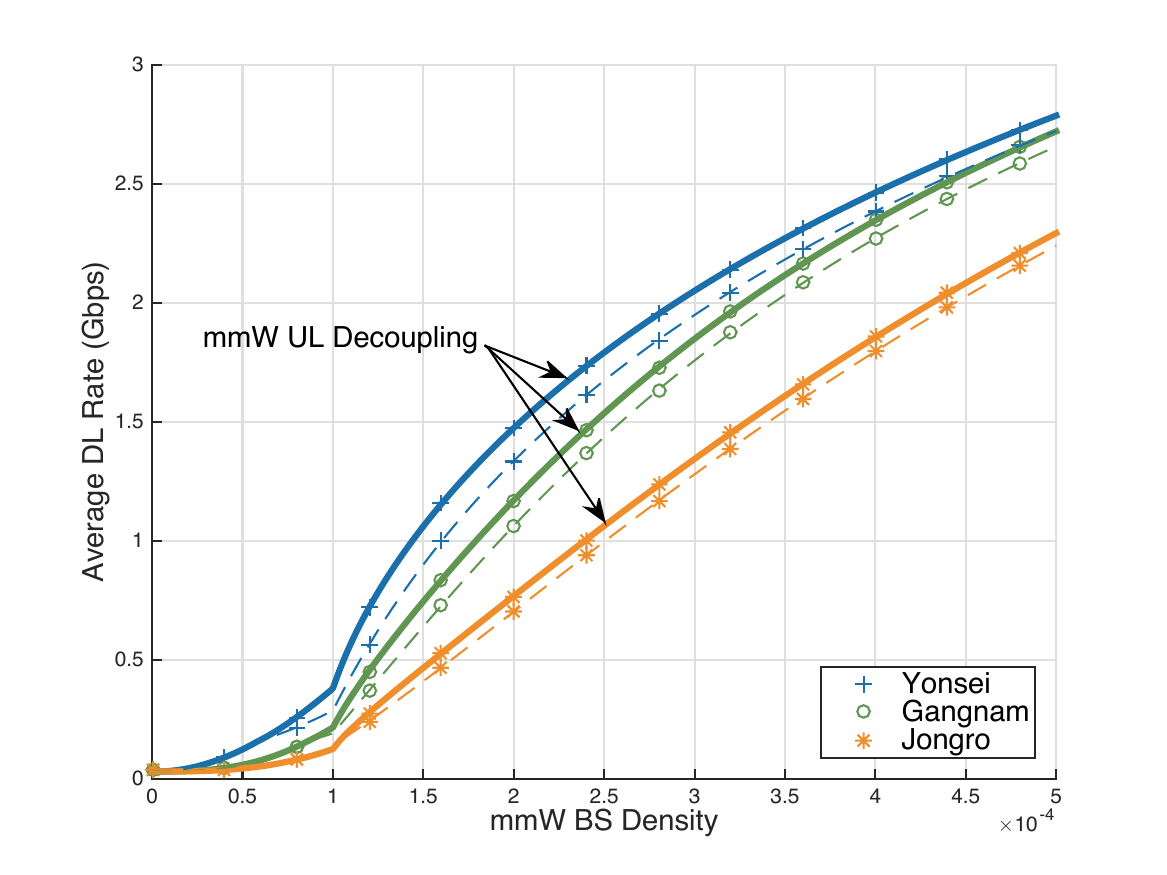} 
\caption{Maximized average DL rate with and without mmW UL decoupling ($W_m = 2$ GHz, $\zeta = 0.25$) } \label{Fig:Rate}
\end{figure}

Fig.~\ref{Fig:Rate} shows the overall DL rates with and without mmW UL decoupling under the proposed mm-$\mu$W resource allocations (Propositions 5 and 7). The DL rate improvements are up to $33$\% in Yonsei and $15$\% in Jongro.

\begin{figure}
\vspace{10pt}\subfigure[ For $W_m = 3$ GHz ]{\includegraphics[width=9cm]{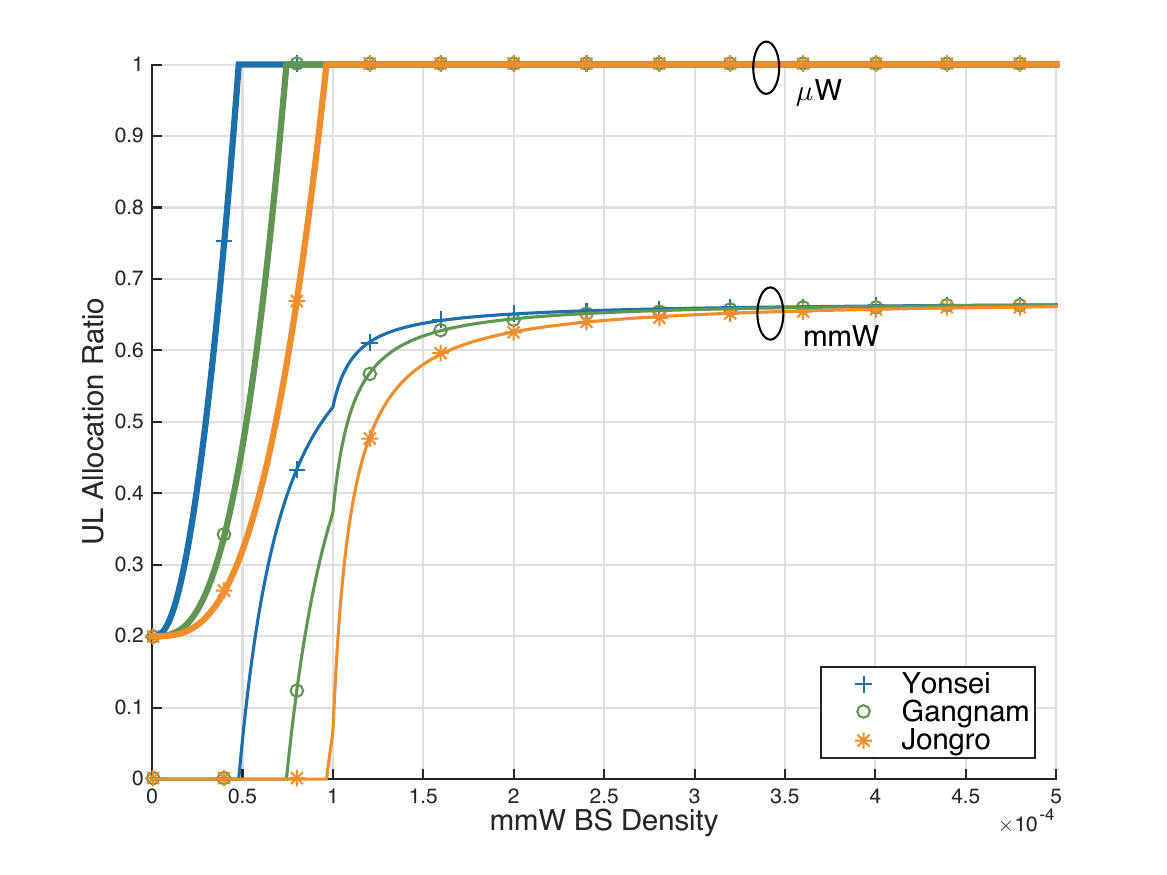}} 
\vspace{10pt}\subfigure[ For $\zeta=0.5$]{\includegraphics[width=9cm]{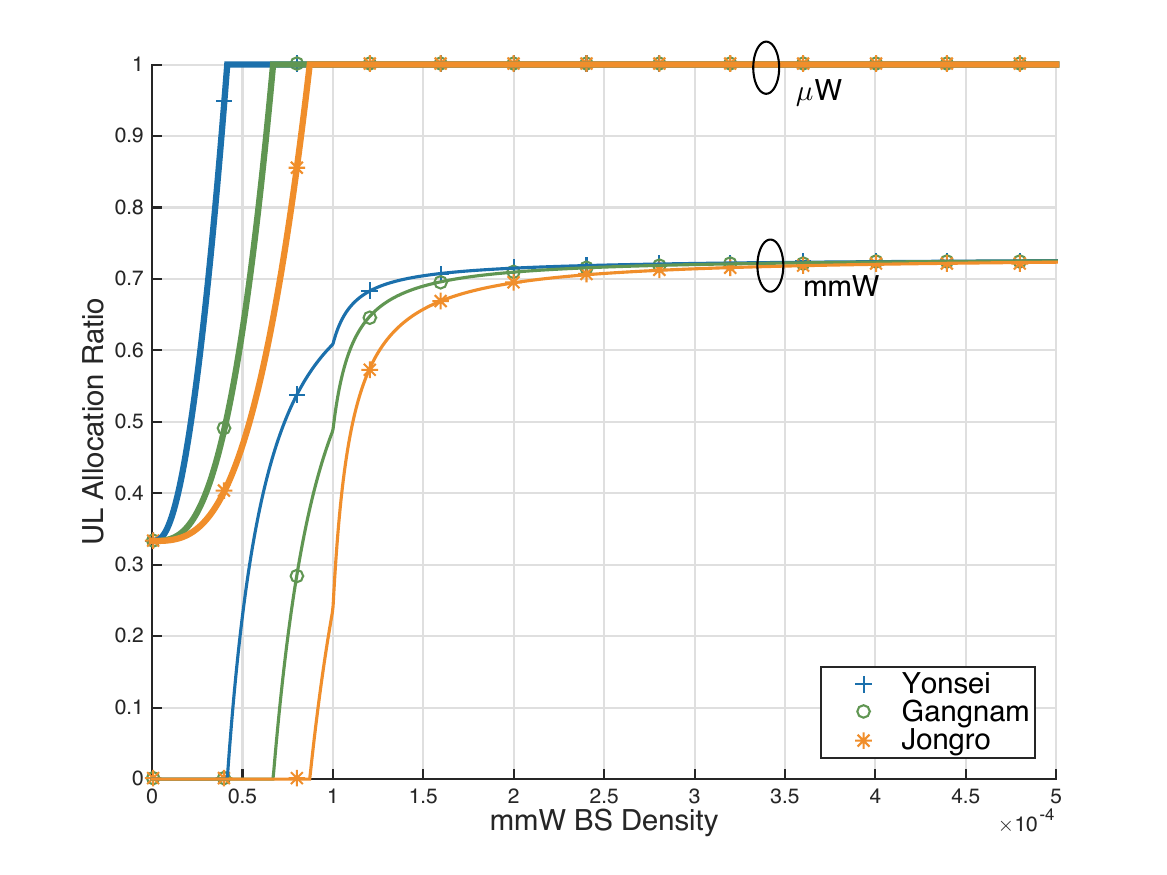}} 
\vspace{10pt}\subfigure[ For DAC power limited systems]{\includegraphics[width=9cm]{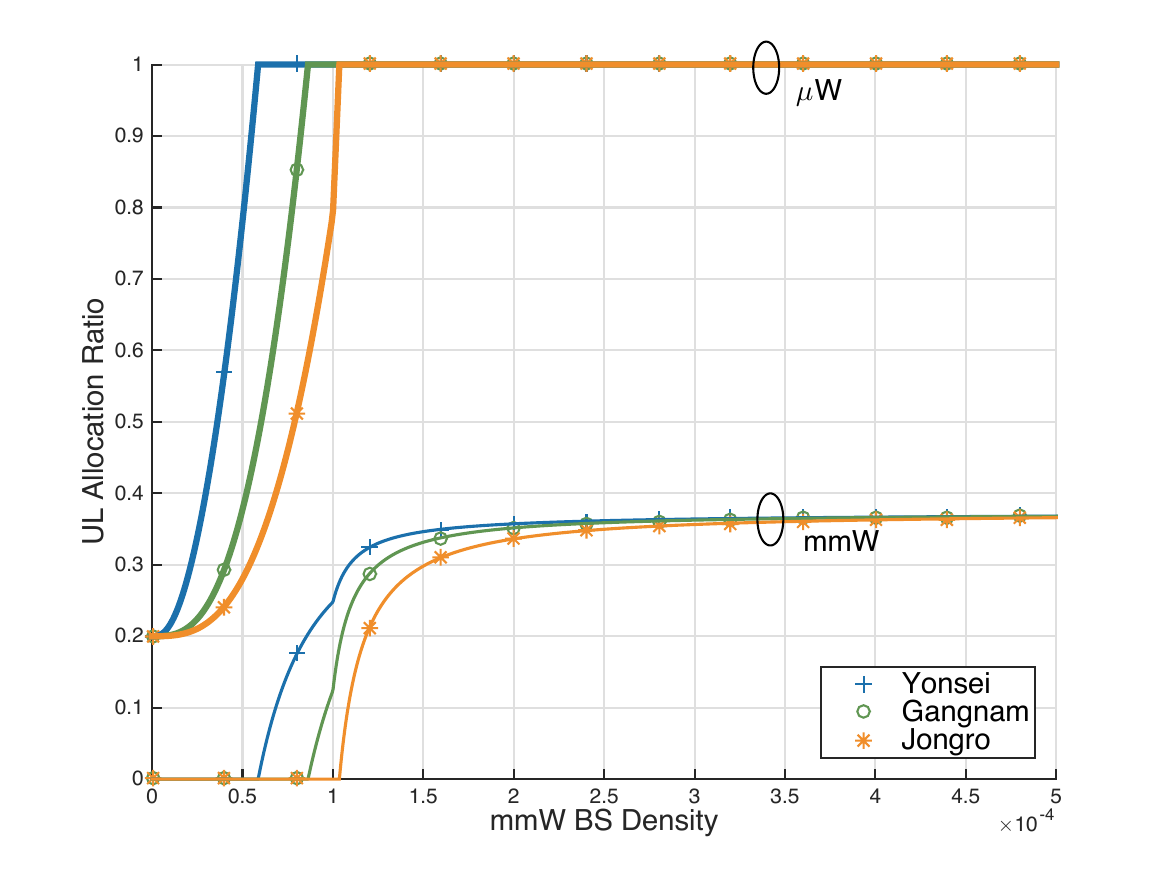}}
\caption{ Optimal mm-$\mu$W UL resource allocations for different environments: (a) mmW bandwidth $W_m$ increased, (b) minimum UL to DL rate ratio $\zeta$ increased, and (c) DAC power limited systems } \label{Fig:Comparison}
\end{figure}

Fig.~\ref{Fig:Comparison} provides the mm-$\mu$W resource allocations with different system parameters from the case in Fig.~\ref{Fig:RscwoDC}. For mmW bandwidth $\WM$ increase to $3$ GHz in Fig.~\ref{Fig:Comparison}-a, mmW UL allocation increases, by $1.37$ times in Yonsei at $\lambdahM=1$ and by $1.2$ times for the maximum mmW UL allocations. 

For minimum UL to DL rate ratio $\zeta$ increase to $50$\% in Fig.~\ref{Fig:Comparison}-b, both mm-$\mu$W UL allocations increase: up to $70$\% $\mu$W allocation increase; $58$\% mmW increase in Yonsei at $\lambdahM=1$ and $31$\% increase for the maximum mmW UL allocations. It is worth noting that the minimum $\mu$W UL allocations depend solely on $\zeta$ by comparing Figs. 7, 8, and 10 (Corollary 2).

For DAC power limited systems in Fig.~\ref{Fig:Comparison}-c, mmW UL allocation in Yonsei at $\lambdahM=1$ decreases by $34$\%, and the maximum mmW UL allocations also decrease by $33$\%, compared to the allocations in Fig. 7. Note that our simulation settings in Fig.~\ref{Fig:RscwoDC} are PAPR outage limited, i.e. PAPR limited mmW UL bandwidth is smaller than the DAC power limited mmW UL bandwidth in \eqref{Eq:mmWLimitedUL}. This may correspond with an OFDMA scenario where a large number of multiple carriers are likely to incur high PAPR. Neglecting the PAPR limited mmW UL bandwidth in \eqref{Eq:mmWLimitedUL} yields DAC power limited systems in Fig.~\ref{Fig:Comparison}-c, which may correspond to an SC-FDMA environment where PAPR outage hardly occurs.


\vspace{-10pt}\section{Conclusion} \label{Sect:Conclusion}
This paper tackles the mmW UL bottleneck due to high PAPR in 5G mm-$\mu$W cellular networks. As its solution, inter-tier DL/UL resource management in concert with mmW UL decoupling is proposed, and its impact is analyzed in a tractable manner. The result indicates that mm-$\mu$W resource allocations have high priorities respectively in DL and UL (Proposition 3). In addition, it shows mmW UL decoupling mitigates the UL bottleneck, thereby increasing DL rate (Proposition 7). The closed-form mm-$\mu$W UDN SEs (Theorems 1 and 2) are key enablers to provide such an intuitive guideline. A real geography based 3D blockage model enhances the practical feasibility of our analysis.

The tractable analysis of this study removes the complicated representations in mm-$\mu$W resource management, yet in return reduces its practical viability. To further validate our result, its follow-up investigation may incorporate more realistic near-field channel behaviors. 3D channel characteristics and diverse path loss attenuations are major candidates.

Further extension could also contemplate the impact of DL transmission outage. Our mm-$\mu$W resource management in the end suggests mmW exclusive DL transmissions from the average DL rate point of view. The blockage vulnerable nature of mmW, however, makes it incapable of supporting ultra-reliable applications such as the tactile internet. The complementary DL use of $\mu$W when encountering blockages can be its possible solution. Analyzing such a microscopic operation in view of the mm-$\mu$W resource allocations is another interesting topic for future work.

\section*{Appendix I. Proof of Lemmas}

\setcounter{subsection}{0}
\subsection{Proof of Lemma 1 (Homogeneous Active BS Approximation)}
Consider the Voronoi cell size distribution \cite{Ferenc:07} where the cell size is normalized by BS density. Restoring BS density $\lambda$ rephrases the cell size probability density distribution as

{\small\begin{align}
f_X(x) &= \frac{3.5^{3.5}}{\Gamma(3.5)} \lambda{}^{4.5} x^{3.5} e^{-3.5 \lambda x}
\end{align}\normalsize}
where X denotes the cell size. The variance of Voronoi cell sizes is then represented as follows.
\small\begin{align}
\text{Var(X)}  = \E X^2 - (\E X)^2 &= \frac{5.5\times 4.5}{3.5^2 \lambda{}^2} - \(\frac{4.5^2}{3.5 \lambda}\)^2\\
& = \frac{4.5}{3.5^2 \lambda{}^2} \label{Eq:CellSizeVar}
\end{align}\normalsize

As $\lambda$ (or $\hat{\lambda}$) goes to infinity, \eqref{Eq:CellSizeVar} approaches zero, indicating location dependency of the activation (or cell size) marks becomes negligible. If users are uniformly distributed, this makes the active BS PPP be regarded as a homogeneous PPP that is independently thinned by $p_a$. Applying Taylor expansion to $p_a$ for $\hat{\lambda} \rightarrow \infty$ yields $p_a = \hat{\lambda}^{-1}$, and thus the resultant homogeneous PPP density $p_a \lambda$ converges to $\lambdaU$. \endproof

\begin{figure*} 
\setcounter{equation}{52}
\small\begin{align}
\gammaMuDL &\geq \log\( 1 +\[\rho_{\mu}{}^{-1}\lambdahMu\]^{\frac{\alphaMu}{2}}\) + \frac{\alphaMu}{2}\bigg[ \rho_{\mu}\lambdahMu^{-1} - \( 1 + \lambdahMu^{-\frac{\alphaMu}{2}}\)^{\frac{2}{\alphaMu}} \underbrace{ \, _2F_1\left(-\frac{2}{\alphaMu},-\frac{2}{\alphaMu};1-\frac{2}{\alphaMu};\frac{1}{1 + \lambdahMu^{\frac{\alphaMu}{2}}} \right)}_{(b)} \bigg] \label{Eq:SEMuLB}\\
\gammaMuDL & \leq \( 1 - \lambdahMu^{-1}\) \int_{t>0} \[ 1- \(\lambdahMu - 1 \)^{-1} \frac{\alphaMu}{\alphaMu + 2} (e^t - 1)^{\frac{2}{\alphaMu}} \]^+ dt \\
&=  \( 1-\lambdahMu^{-1}\) \log\( 1 + \[\(\lambdahMu - 1\)\(1 + \frac{2}{\alphaMu}\)^\frac{\alphaMu}{2} \]\) - \frac{\lambdahMu}{\( \lambdahMu - 1\)^2} \lb  \frac{2}{\alphaMu}+ \(\frac{2}{\alphaMu}\)^2\rb^{-1}\nn\\
&\hspace{8pt}\times \bigg[  \rho_{\mu}  - \lb 1 + \[ \(1 + \frac{2}{\alphaMu} \)\(\lambdahMu - 1 \)^{\frac{\alphaMu}{2}} \]^{\frac{2}{\alphaMu}} \rb  \underbrace{ \, _2F_1\left(-\frac{2}{\alphaMu},-\frac{2}{\alphaMu};1-\frac{2}{\alphaMu};\lb 1 + \[ \(1 + \frac{2}{\alphaMu} \)\(\lambdahMu - 1 \)^{\frac{\alphaMu}{2}} \]^{\frac{2}{\alphaMu}} \rb^{-1} \right)}_{(c)} \bigg] \label{Eq:SEMuUB}
\end{align}\normalsize
\hrulefill
\end{figure*}

\setcounter{equation}{36}
\subsection{Proof of Lemma 2 (UDN Approximation)}
Appying the nearest BS distance distribution \cite{HaenggiSG},
\small\begin{align}
\E_{R}\[ \exp\( -\lambdaU \pi r^2 I \)\]  &=  2  \lambda \pi \int_{r>0} r \exp\lb -\lambda \pi r^2 - \lambdaU \pi r^2 I   \rb dr \\
&= \[ 1 + \hat{\lambda}{}^{-1}I \]^{-1}.
\end{align}\normalsize
Exploiting Taylor expansion while preventing negative values completes the proof. \endproof

\subsection{Proof of Lemma 3 ($\mu$W Interference Constant Bounds)}
The proof of the upper bound is trivial when recalling $\rhoMu$'s integration range that is larger than the range of $\rhoMut$. For the lower bound, applying Taylor expansion at the integrand in $\rhoMut$ yields

\small\begin{align}
\rhoMut &\geq \int_{u>(e^t-1)^{-\frac{2}{\alphaMu}}} \(1 - u^{\frac{\alphaMu}{2}}\)^+du \\
&= \(1 + \frac{2}{\alphaMu} \)^{-1}\[1 + (e^t - 1)^{-\( 1 + \frac{2}{\alphaMu}\)} \] -(e^t - 1)^{-\frac{2}{\alphaMu}} \\
&\geq  \(1 + \frac{2}{\alphaMu} \)^{-1} - (e^t - 1)^{-\frac{2}{\alphaMu}} \label{Eq:PfProp1UB_rhot}
\end{align}\normalsize
from which the result follows. \endproof

\subsection{Proof of Lemma 4 (Directional Interference Thinning)}
Consider the DL aggregate interference at a typical user.
\small\begin{align}
I_{\Sigma_{\text{m.d}}} &:=  \sum_{i\in \PhiActM} \Theta_i g_i {r_i}^{-\alphaM} \mathds{1}_{\RL}(r_i)  \\
&\overset{(a)}{\approx} \sum_{i\in \Phi_{\text{m}}\( p_\mu \lambdaM\)} \Theta_i g_i {r_i}^{-\alphaM} \mathds{1}_{\RL}(r_i) \\
& \overset{(b)}{=} \sum_{i \in \Phi_{\text{m}}\( p_\text{m} \lambdaM\(\frac{\theta}{2\pi}\)^{\frac{2}{\alphaM}}\)} g_i {r_i}^{-\alphaM} \mathds{1}_{\RL}(r_i)
\end{align}\normalsize
where $(a)$ follows from Lemma 1 and $(b)$ from: (i) $\E \Theta_i  \frac{\theta}{2 \pi}$ due to uniformly distributed users and (ii) mapping theorem, also known as displacement theorem \cite{HaenggiSG}. The UL aggregate interference representation follows the same procedure. \endproof

\subsection{Proof of Lemma 5 (mmW Interference Constant Bounds)}
The upper bound is straightforwardly derived when comparing the integration ranges of $\rhoMt$ and $\rhoM$. For the lower bound, applying Taylor expansion at the integrand of $\rhoMt$ yields
\small\begin{align}
\rhoMt &\geq \int_{ (e^t-1)^{-\frac{2}{\alphaM}}}^{\min\lb 1, \(\frac{\RL}{r}\)^2 (e^t-1)^{-\frac{2}{\alphaM}}\rb} \( 1 - u^{\frac{\alphaM}{2}}\) du. \label{Eq:PfLemma5}
\end{align}\normalsize

Assume the upper limit of the integration in \eqref{Eq:PfLemma5} becomes $1$, yielding
\small\begin{align}
\rhoMt &\geq  1 -\(e^t-1\)^{-\frac{2}{\alphaM}} - \(1 + \frac{\alphaM}{2}\)^{-1} \[ 1-  \(e^t-1\)^{-\frac{2} {\alphaM}-1} \] \\
&\geq 1 -\(e^t-1\)^{-\frac{2}{\alphaM}} - \(1 + \frac{\alphaM}{2}\)^{-1}. 
\end{align}\normalsize

Consider the integration upper limit in \eqref{Eq:PfLemma5} becomes $1$ with probability no less than $1-\varepsilon$ for a constant $\varepsilon>0$, i.e.
\small\begin{align}
\Pr\( R \leq \RL (e^t-1)^{-\frac{1}{\alphaM}}  \)  &=  1- e^{-\lambdaM \pi \RL^2 (e^t-1)^{-\frac{2}{\alphaM}}} \geq 1-\varepsilon.
\end{align}\normalsize
It is rephrased by
\small\begin{align}
t\leq \log\[1 + \(\frac{\lambda_m \pi \RL^2}{\log\varepsilon^{-1}} \)^{\frac{\alphaM}{2}}\]. \label{Eq:PfLemma5_2}
\end{align}\normalsize

For $\lambdaM \rightarrow \infty$, the inequality \eqref{Eq:PfLemma5_2} always holds for all $t$, resulting in the first argument of the lemma. For a UDN, when applying Lemma 2 to \eqref{Eq:PfProp1ExactPre}, the value $t$ is upper bounded by $\log\[1 + \(\frac{\lambda_m }{\rhoM \lambdaU} \)^{\frac{\alphaM}{2}}\]$. Making this upper bound belong to the inequality in \eqref{Eq:PfLemma5_2} is identical to satisfying \eqref{Eq:PfLemma5_2} for all $t$. This leads to $\varepsilon \geq e^{-\rhoM \lambdaU \pi \RL^2}$. Taking the minimum value of $\varepsilon$ completes the proof.
\endproof

\begin{figure*}\setcounter{equation}{59}
\small\begin{align}
\gammaMDL & \leq \( 1 - \frac{1}{\lambdahM}\) \int_{t>0} \(1 - e^{-\lambdaM \pi \RL^2 \lb 1 + \frac{1}{\lambdahM\(1+\frac{2}{\alphaM}\)} \[N^{-\frac{1}{2}} (e^t-1)\]^{\frac{2} {\alphaM}} \rb}\) \l\{ 1- \[\(\lambdahM - 1 \)\(1 + \frac{2}{\alphaM}\) \]^{-1}\[N^{-\frac{1}{2}}(e^t - 1)  \]^{\frac{2}{\alphaM}} \r\}^+ \hspace{-5pt} dt \\
&\overset{(a)}{\leq} \( 1 - \frac{1}{\lambdahM}\) \int_{t>0} \pLt \[ 1- \[\(\lambdahM - 1 \)\(1 + \frac{2}{\alphaM}\) \]^{-1}\[N^{-\frac{1}{2}}(e^t - 1)  \]^{\frac{2}{\alphaM}} \]^+ dt \label{Eq:PfProp3UB}
\end{align}\normalsize
\hrulefill
\end{figure*}

\section*{Appendix II. Proofs of Propositions}
\setcounter{subsection}{0}

\subsection{Proof of Proposition 1 ($\mu$W DL/UL SE Bounds)} \label{Sect:PfProp1}
According to Lemma 1, interferer PPPs for DL $\PhiActMu$ and for UL $\PhiActMu$ in a UDN can be regarded as independently thinned homogeneous PPPs respectively from $\Phi_\mu$ and $\Phi_u$. For DL $\mu$W UDN SE at a typical user, the interferer density is then given as $p_a \lambdaMu{P_{\mu.d}}^{\frac{2}{\alphaMu}} \approx \lambdaU{P_{\mu.d}}^{\frac{2}{\alphaMu}}$ where the approximation results from Taylor expansion for a UDN. Note that ${P_{\mu.d}}^{\frac{2}{\alphaMu}}$ follows from mapping theorem \cite{HaenggiSG} that interchanges $P$ times power increase with $P^{\frac{2}{\alpha}}$ times density increase. Similarly, the density of the desired BS, denoting the associated BS transmitting the desired DL signal, is represented by $\lambdaMu {P_{\mu.d}}^{\frac{2}{\alphaMu}}$ thanks to mapping theorem. 

Starting from this setting, applying the equation (16) in \cite{Andrews:2011bg} with minor modification yields the $\mu$W DL UDN SE $\gammaMuDL$ as

\setcounter{equation}{49}{\small\begin{align}
\gammaMuDL 
& = \int_{t>0} \E_{\PhiActMu}\[ \Pr\lb \log\( 1 + \SIR_{\mu.d}  \) \geq t \rb \] dt \\
&=  \int_{t>0} \E_R \[  \exp\( - \pi r^2 \rho_\mu^{(t)} \lambdaU {P_{\mu.d}}^{\frac{2}{\alphaMu}}    (e^t - 1)^{\frac{2}{\alphaMu} }     \) \] dt \label{Eq:PfProp1_Pre}\\
&\overset{(a)}{\approx}  \int_{t>0} \[ 1- \rho_{\mu}^{(t)} \lambdahMu^{-1} (e^t - 1)^{\frac{2}{\alphaMu}} \]^+ dt \label{Eq:PfProp1}
\end{align}\normalsize}
where $(a)$ follows from Lemma 2 and the probability density function (pdf) of the BS-to-user association distance $R$: $f_R(r) = 2 \pi r \lambdaMu {P_{\mu.d}}^{\frac{2}{\alphaMu}}  e^{-\pi r^2 \lambdaMu {P_{\mu.d}}^{\frac{2}{\alphaMu}} }$.

Applying the upper bound of $\rho_\mu^{(t)}$ in Lemma 3 leads to the lower bound of $\gammaMuDL$ as \eqref{Eq:SEMuLB} at the top of this page, where Gaussian hypergeometric function $\, _2F_1(a, b; c; z) : = \sum _{k=0}^{\infty } \frac{z^k a^{(k)} b^{(k)} }{k! c^{(k)}}$, and $x^{(k)}$ rising factorial. The function $(b)$ monotonically decreases with $\lambdahMu$ for all $\alphaMu>2$, having its maximum $\rho_{\mu}$ and minimum $1$ respectively at $\lambdahMu=0$ and $\lambdahMu\rightarrow\infty$ \cite{AbramowitzBook:HandbookFunctions:1995}. Considering $\lambdahMu\gg 1$ yields the desired DL $\mu$W UDN SE lower bound.

In a similar manner, applying the lower bound of $\rho_\mu^{(t)}$ in Lemma 3 results in the upper bound of $\gammaMuDL$ as \eqref{Eq:SEMuUB} at the top of the page. The function $(c)$ monotonically decreases with $\lambdahMu$ for all $\lambdahMu \geq 1$ and $\alphaMu>2$, having its maximum $\rho_{\mu}$ and mininum $1$ respectively at $\lambdahMu = 1$ and $\lambdahMu\rightarrow \infty$ \cite{AbramowitzBook:HandbookFunctions:1995}. Considering $\lambdahMu \gg 1$ leads to the desired DL $\mu$W UDN SE upper bound.

For UL $\mu$W UDN SE at a typical BS, the interferer and the desired BS densities are respectively represented as $ p_s \lambdaU {P_{\mu.u}}^{\frac{2}{\alphaMu}}\approx \lambdaU {P_{\mu.u}}^{\frac{2}{\alphaMu}}$ and $\lambdaMu {P_{\mu.u}}^{\frac{2}{\alphaMu}}$ due to mapping theorem \cite{HaenggiSG} and Taylor expansion. When noticing the ${P_{\mu.d}}^{\frac{2}{\alphaMu}}$'s of the interferer and the desired BS are cancelled out in \eqref{Eq:PfProp1}, such an UL and DL transmitting power difference in the interferer and the desired BS densities do not affect the SE calculation afterwards. As a consequence, UL $\mu$W UDN SE shares the same lower and upper bound with the DL case, completing the proof.
\endproof

\begin{figure*}
\centering
 	\subfigure[if $W_{m.u}\gammaMUL < W_m \gammaM$ (always holds in the cases without mmW UL decoupling)]{\includegraphics[width=11cm]{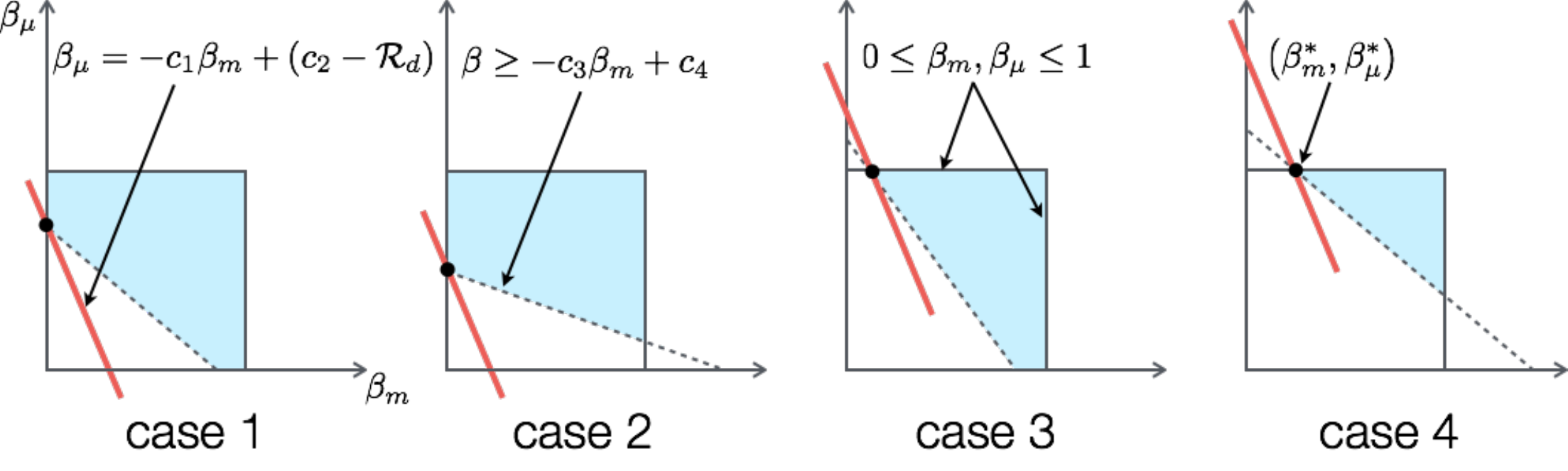}} \label{Fig:RscwoDCPf} \hspace{30pt}
	\subfigure[if $W_{m.u}\gammaMUL \geq W_m \gammaM$]{\includegraphics[width=5.25cm]{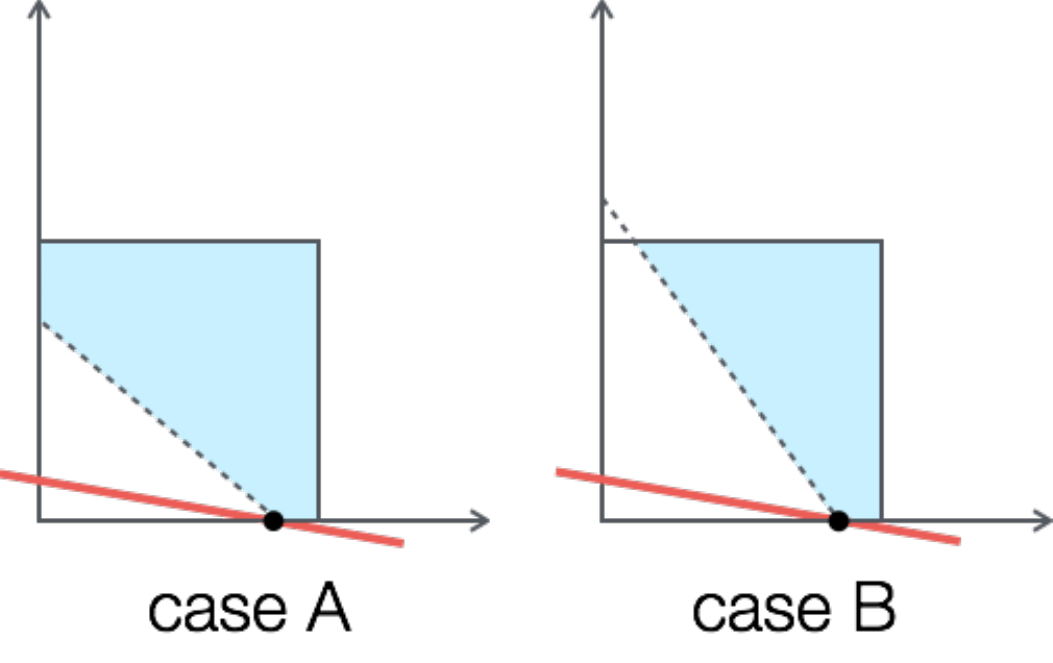}} \label{Fig:RscwDCPf}
	\caption{Illustrations of the problem \textsf{P1} in Section \ref{Sect:RscProblem} for different settings in \eqref{Eq:ConstULQoS2} without and with mmW UL decoupling.} \label{Fig:RscPf}
\end{figure*}

\subsection{Proof of Proposition 2 (mmW DL/UL SE Bounds)}
In a similar procedure to the proof of Proposition 1, the interferer PPP for DL $\PhiActM$ is approximated by a homogeneous PPP with density $\lambdaU\( N^{-\frac{1}{2}}P_{m.d} \)^{\frac{2}{\alphaM}}$, and for UL $\PhiActU$ by also a homogeneous PPP with density $\lambdaU \( N^{-\frac{1}{2}} P_{m.u}\)^{\frac{2}{\alphaMu}}$ according to Lemmas 1 and 4 with exploiting Taylor expansion under a UDN regime. For DL mmW SE, the desired BS density is represented as $\lambdaM{P_{m.d}}^{\frac{2}{\alphaM}}$, yielding

\setcounter{equation}{55}{\small\begin{align}
\gammaMDL 
&= \int_{t>0} \E_{\PhiActM}\[ \Pr\lb \log\( 1 + \SIR_{m.d}  \) \geq t \rb \] dt \\
&\hspace{-15pt}= \int_{t>0} \E_R \[  \exp\( - \pi r^2 \rho_\mu^{(t)} \lambdaU \(N^{-\frac{1}{2}} P_{m.d}\)^{\frac{\alphaMu}{2}}    (e^t - 1)^{\frac{2}{\alphaM} }    \) \] dt \\
&\hspace{-15pt}=2 \pi \lambdaM {P_{m.d}}^{\frac{2}{\alphaM}} \int_{t>0} \int_{0}^{\RL} r \exp \Bigg\{  - \pi r^2 \lambdaM  {P_{m.d}}^{\frac{2}{\alphaM}} \nn\\
&- \rho_m^{(t)} \lambdaU {P_{\mu.d}}^{\frac{2}{\alphaMu}} \pi r^2 \[ N^{-\frac{1}{2}}\( e^t-1\) \]^{\frac{2}{\alphaM}}  \Bigg\} drdt \label{Eq:PfProp2}
\end{align}\normalsize}
where the last result comes from the pdf of $R$: $f_R(r) = 2 \pi r \lambdaM {P_{m.d}}^{\frac{2}{\alphaM}} e^{-\pi r^2\lambdaM {P_{m.d}}^{\frac{2}{\alphaM}} }$.

Applying the upper bound in Lemma 5 results in the lower bound of DL mmW SE as below.

\small\begin{align}
\gammaMDL &\geq \int_{t>0} \pLt \[ 1 + \lambdahM^{-1}\rhoM \[ N^{-\frac{1}{2}} (e^t-1)\]^{\frac{2}{\alphaM}}\]^{-1} dt 
\end{align}\normalsize

Next, exploiting the lower bound in Lemma 5 yields the upper bound of DL mmW SE as \eqref{Eq:PfProp3UB} at the top, where $(a)$ follows from $\rhoM \geq 1-\(1 + \frac{\alphaM}{2} \)^{-1}$. Considering $\lambdahMu \gg 1$ yields the desired DL mmW SE upper bound. 

As in the case of the proof of Proposition 1, the transmitting power effects at the desired signal and interference are mutually cancelled out after \eqref{Eq:PfProp2}. It makes the lower and upper bounds of DL and UL be identical, finalizing the proof. \endproof

\subsection{Proof of Proposition 3 (mmW Sparse SE)}
Consider a noise-limited mmW DL network. For a given $t$, $\SNR$ coverage probability is represented as below.
\small\begin{align}
\Pr\(\SNR>t\) &= \Pr\(\log\[ 1 + \frac{G_N P_{m.d}  g R^{-\alpha_\text{m}}}{\sigma^2}\]>t \mid R \leq \RL \) \nonumber\\
&\quad \times \Pr\(R \leq \RL\)\\
&= \int_0^{\RL} e^{-\frac{\sigma^2 r^{\alphaM}\(e^t-1\)}{N P_{m.d}}}f_R(r)dr \label{Eq:PfsparseCov}
\end{align}\normalsize
Taking an integration over $t$ with the order change of integrations yields the desired result, i.e.
\small\begin{align}
\gammaMDL &= \lambdahM\int_0^\infty \int_0^{\RL} e^{-\frac{\sigma^2 r^{\alphaM}\(e^t-1\)}{N P_{m.d}}}f_R(r)dr dt\\
&= \lambdahM \int_0^{\RL} \[\int_0^\infty e^{-\frac{\sigma^2 r^{\alphaM}\(e^t-1\)}{N P_{m.d}}}dt\] f_R(r)dr 
\end{align}\normalsize
where $\lambdahM$ in front of integrations indicates a typical user's selection probability by its associated mmW BS under a uniformly random scheduler. The derivation for mmW UL SE follows the same procedure above, concluding the proof.\endproof

\subsection{Proof of Proposition 4 (Optimal mm-$\mu$W UL Resource Allocations)} \label{Sect:PfProp3}
Consider $\RDL$ in \eqref{Eq:RateDL} that is rephrased by
\setcounter{equation}{61}\small\begin{align}
\beta_\mu &= -c_1 \beta_m + (c_2 - \RDL) \label{Eq:RateDL2}
\end{align}\normalsize
where $c_1:= W_m \gamma_m / (W_\mu \gamma_\mu)$ and $c_2:= W_m \gamma_m + W_\mu \gamma_\mu $. Note that $c_2\geq \RDL$ since $c_2$ is the maximum DL rate achieved when the entire mm-$\mu$W resource are dedicated to DL. The minimum UL/DL rate ratio in \eqref{Eq:ConstULQoS} is also represented as
\small\begin{align}
\beta_\mu &\geq -c_3 \beta_m + c_4 \label{Eq:ConstULQoS2}
\end{align}\normalsize
where $c_3:= \[ W_{m.u}\gamma_{m.u} + \zeta W_m \gamma_m \]\[ (1 + \zeta)W_\mu \gamma_\mu\]$ and $c_4:=\(1 + \zeta^{-1}\)^{-1}\(1 + \frac{W_m \gamma_m}{W_\mu \gamma_\mu}\)$.

The given \textsf{P1} in Section \ref{Sect:RscProblem} is then equivalent to a the following graphical problem on the $\beta_m$-$\beta_\mu$ plane: maximizing the $\beta_\mu$-intercept of \eqref{Eq:RateDL2} within the feasible region formed by \eqref{Eq:ConstULMax} and \eqref{Eq:ConstULQoS2}. Solving this problem requires to consider the four possible cases for different settings in \eqref{Eq:ConstULQoS2} as visualized in Fig.~\ref{Fig:RscPf}-a. 

The cases 1 and 2 have the identical solution $(\beta_m^*,\beta_\mu^*) =\(0, \[1 + \frac{W_m \gammaM}{W_\mu \gammaMu}\]\[1 + \zeta^{-1}\]^{-1} \)$, and share the same condition $\zeta \leq W_\mu \gammaMu /\(W_m \gammaM \)$. Similarly, the cases 3 and 4 having the identical solution $(\beta_m^*,\beta_\mu^*) =\( \[\zeta - \frac{W_\mu \gammaMu}{W_m \gammaM} \]\[ \zeta + \frac{W_{m.u}\gammaMUL}{W_m \gammaM} \]^{-1} ,1\)$ share the same condition $\zeta > \frac{W_\mu \gammaMu}{W_m \gammaM}$. Applying Theorems 1 and 2 to these results conclude the proof.\endproof

\subsection{Proof of Proposition 5 (Optimal mm-$\mu$W UL Resource Allocations with mmW UL Decoupling)} \label{Sect:PfProp5}
Consider $W_{m.u}\gammaMUL < W_m \gammaM$ that always holds in the cases without mmW UL decoupling. If this holds, the solutions with mmW UL decoupling are identical to the results in Proposition 3. In addition, it is necessary to consider $W_{m.u}\gammaMUL \geq W_m \gammaM$ under the cases with mmW UL decoupling. This corresponds with the scenarios where the mmW UL receptible BS densification overcomes the mmW UL bandwidth limitation. For such scenarios, the possible cases A and B are visualized in Fig.~\ref{Fig:RscPf}-b. These cases have the identical solution $(\beta_m^*,\beta_\mu^*) = \( \[ 1 + \frac{W_\mu \gamma_\mu}{W_m \gammaM}\] \[1 + \frac{W_{m.u}\gammaMUL}{\zeta W_m \gammaM}\]^{-1},0 \)$, and share the same condition $\zeta \leq W_{m.u}\gamma_{m.u}/\( W_\mu \gammaMu\)$. Note that this condition is always true due to the assumptions \textsf{A1} and \textsf{A2} in Section \ref{Sect:RscProblem}, and the opposite condition is thus false. Combining the solutions with Theorems 1 and 2 leads to the desired result.\endproof

\section*{Acknowledgement}\small
The authors are grateful to June Hwang and Minho Kim in Samsung Electronics for helpful feedback on the mmW overlaid cellular network architecture.

\bibliographystyle{ieeetr}

\begin{thebibliography}{10}

\bibitem{SamsungmmWave:11}
Z.~Pi and F.~Khan, ``{An Introduction to Milimeter-Wave Mobile Broadband
  Systems},'' {\em IEEE Communications Magazine}, vol.~49, no.~6, pp.~101--107,
  2011.

\bibitem{SamsungSarnoff11}
Z.~Pi and F.~Khan, ``{System Design and Network Architecture for a
  Millimeter-wave Mobile Broadband (MMB) System},'' {\em in Proc. IEEE Sarnoff
  Symposium, Princeton, NJ, United States}, 2011.

\bibitem{SamsungGC:13}
T.~Kim, J.~Park, J.-Y. Seol, S.~Jeong, J.~Cho, and W.~Roh, ``{Tens of Gbps
  Support with mmWave Beamforming Systems for Next Generation
  Communications},'' {\em Proc. IEEE Global Communications Conference
  (GLOBECOM)}, pp.~3685--3690, 2013.

\bibitem{Ericsson5G:13}
Ericsson, ``{5G Radio Access},'' {\em Ericsson Review}, February 2015.

\bibitem{Rappaport5G:13}
T.~S. Rappaport, S.~Sun, R.~Mayzus, H.~Zhao, Y.~Azar, K.~Wang, G.~N. Wong,
  J.~K. Schulz, M.~Sammi, and F.~Gutierrez, ``{Milimeter Wave Mobile
  Communications for 5G Cellular: It Will Work!},'' {\em IEEE Access}, vol.~1,
  pp.~335--349, 2013.

\bibitem{Rappaport:14}
S.~Ragan, T.~S. Rappaport, and E.~Erkip, ``{Milimeter-Wave Cellular Wireless
  Networks: Potentials and Challenges},'' {\em Proceedings of the IEEE},
  vol.~102, no.~3, pp.~366--385, 2014.

\bibitem{Andrews5G:14}
J.~G. Andrews, S.~Buzzi, W.~Choi, S.~Hanly, A.~Lozano, A.~C.~K. Soong, and
  J.~C. Zhang, ``{What Will 5G Be?},'' {\em IEEE Journal on Selected Areas in
  Communications}, vol.~PP, no.~99, 2014.

\bibitem{Holistic13}
I.~Hwang, B.~Song, and S.~S. Soliman, ``{A Holistic View on Hyper-Dense
  Heterogeneous and Small Cell Networks},'' {\em IEEE Communications Magazine},
  vol.~51, no.~6, pp.~20--27, 2013.

\bibitem{Zander13}
J.~Zander and P.~M{\"a}h{\"o}nen, ``{Riding the Data Tsunami in the Cloud:
  Myths and Challenges in Future Wireless Access},'' {\em IEEE Communications
  Magazine}, vol.~51, no.~3, pp.~145--151, 2013.

\bibitem{Qualcomm:14}
Qualcomm, ``{Hyper-Dense Small Cell Deployment Trial in NASCAR Environment},''
  April 2014.

\bibitem{mmWUDN:15}
R.~Baldmair, T.~Irnich, K.~Balachandran, E.~Dahlman, G.~Mildh, Y.~Sel{\'e}n,
  S.~Parkvall, M.~Meyer, and A.~Osseiran, ``{Ultra-Dense Networks in
  Millimeter-Wave Frequencies},'' {\em IEEE Communications Magazine}, vol.~53,
  no.~1, pp.~202--208, 2015.

\bibitem{Heath:13}
T.~Bai, R.~Vaze, and R.~W. Heath, Jr, ``{Analysis of Blockage Effects on Urban
  Cellular Networks},'' {\em IEEE Transactions on Wireless Communications},
  vol.~13, no.~9, pp.~5070--5083, 2014.

\bibitem{Heath:14}
T.~Bai and R.~W. Heath, Jr, ``{Coverage and Rate Analysis for Millimeter Wave
  Cellular Networks},'' {\em available at: http://arxiv.org/abs/1402.6430.}

\bibitem{Yu2011}
S.~M. Yu and S.-L. Kim, ``{Downlink Capacity and Base Station Density in
  Cellular Networks},'' {\em Proc. IEEE WiOpt Workshop on Spatial Stochastic
  Models for Wireless Networks (SpaSWiN 2013)}, May 2013.

\bibitem{SLeeKHuang12}
S.~Lee and K.~Huang, ``{Coverage and Economy of Cellular Networks with Many
  Base Stations},'' {\em IEEE Communications Letters}, vol.~16, no.~7,
  pp.~1038--1040, 2012.

\bibitem{Alexiou13}
A.~G. Gotsis and A.~Alexiou, ``{On Coordinating Ultra-Dense Wireless Access
  Networks: Optimization Modeling, Algorithms and Insights},'' {\em available
  at: http://arxiv.org/pdf/1312.1577v1.pdf}.

\bibitem{JHPark:14}
J.~Park, S.-L. Kim, and J.~Zander, ``{Asymptotic Behavior of Ultra-Dense
  Cellular Networks and Its Economic Impact},'' {\em in Proc. IEEE Global
  Communications Conference (GLOBECOM), Austin, TX, United Sates}, December
  2014.

\bibitem{JHParkAPWCS:15}
J.~Park, S.-L. Kim, and J.~Zander, ``{Resource Management and Cell Planning in
  Millimeter-Wave Overlaid Ultra-Dense Cellular Networks},'' {\em in Proc. the
  12th IEEE VTS Asia Pacific Wireless Communications Symposium (APWCS),
  Singapore}, August 2015.

\bibitem{JHParkGC:15}
J.~Park, S.-L. Kim, and J.~Zander, ``{Tractable Resource Management in
  Millimeter-Wave Overlaid Ultra-Dense Cellular Networks},'' {\em available at:
  http://arxiv.org/abs/1507.04658.}

\bibitem{TowardUDN:15}
D.~L{\'o}pez-P{\'e}rez, M.~Ding, H.~Claussen, and A.~H. Jafari, ``{Towards 1
  Gbps/UE in Cellular Systems -- Understanding Ultra-Dense Small Cell
  Deployments},'' {\em available at: http://arxiv.org/abs/1503.03912.}

\bibitem{NCT2013}
C.~Hoymann, D.~Larsson, H.~Koorapaty, and J.-F. Cheng, ``{A Lean Carrier for
  LTE},'' {\em IEEE Communications Magazine}, vol.~51, no.~2, pp.~74--80, 2013.

\bibitem{EricssonVTC:15}
M.~Thurfjell, M.~Ericson, and P.~d. Bruin, ``{Network Densification Impact on
  System Capacity},'' {\em in Proc. IEEE Vehicular Technology Conference (VTC),
  Spring, Glasgow, Scotland}, May 2015.

\bibitem{LOSNLOS:15}
C.~Galiotto, N.~K. Pratas, L.~Doyle, and N.~Marchetti, ``{Effect of LOS/NLOS
  Propagation on Ultra-Dense Networks},'' {\em available at:
  http://arxiv.org/abs/1507.01757.}

\bibitem{3GPPMultSlope:15}
M.~Ding, D.~L{\'o}pez-P{\'e}rez, G.~Mao, P.~Wang, and Z.~Lin, ``{Will the Area
  Spectral Efficiency Monotonically Grow as Small Cells Go Dense?},'' {\em
  available at: http://arxiv.org/abs/1505.01920.}

\bibitem{MultSlope:14}
X.~Zhang and J.~G. Andrews, ``{Downlink Cellular Network Analysis with
  Multi-slope Path Loss Models},'' {\em available at:
  http://arxiv.org/abs/1408.0549.}

\bibitem{AndrewsUDN:15}
J.~G. Andrews, X.~Zhang, G.~D. Durgin, and A.~K. Gupta, ``{Are We Approaching
  the Fundamental Limits of Wireless Network Densification?},'' {\em available
  at: http://arxiv.org/abs/1512.00413.}

\bibitem{StoyanBook:StochasticGeometry:1995}
D.~Stoyan, K.~W. S., and J.~Mecke, {\em {Stochastic Gemoetry and its
  Applications}}.
\newblock Wiley, 2nd~ed., 1995.

\bibitem{Rel12Beyond:13}
D.~Astely, E.~Dahlman, G.~Fodor, S.~Parkvall, and J.~Sachs, ``{LTE Release 12
  and Beyond},'' {\em IEEE Communications Magazine}, vol.~51, no.~7,
  pp.~154--160, 2013.

\bibitem{Singh:14}
S.~Singh, M.~N. Kulkarni, A.~Ghosh, and J.~G. Andrews, ``{Tractable Model for
  Rate in Self-Backhauled Millimeter Wave Cellular Networks},'' {\em available
  at: http://arxiv.org/abs/1407.5537.}

\bibitem{Kulkarni:14}
M.~N. Kulkarni, S.~Singh, and J.~G. Andrews, ``{Coverage and Rate Trends in
  Dense Urban mmWave Cellular Networks},'' {\em in Proc. IEEE Global
  Communications Conference (GLOBECOM), Austin, TX, United Sates}, December
  2014.

\bibitem{SinghDecouple:14}
S.~Singh, X.~Zhang, and J.~G. Andrews, ``{Joint Rate and SINR Coverage Analysis
  for Decoupled Uplink-Downlink Biased Cell Associations in HetNets},'' {\em
  available at: http://arxiv.org/abs/1412.1898.}

\bibitem{HeathWearable:15}
K.~Venugopal, M.~C. Valenti, and R.~W. Heath, Jr, ``{Interference in
  Finite-Sized Highly Dense Millimeter Wave Networks},'' {\em in Proc.
  Information Theory and Applications (ITA), San Diego, CA, United States},
  February 2015.

\bibitem{JLee:14}
J.~Lee and C.~Tepedelenlio{\u g}lu, ``{Stochastic Ordering of Interference in
  Large-Scale Wireless Networks},'' {\em IEEE Transactions on Signal
  Processing}, vol.~62, no.~3, pp.~729--740, 2014.

\bibitem{SamsungDAC:14}
J.~Karjalainen, M.~Nekovee, H.~Benn, W.~Kim, J.~Park, and S.~Hwang,
  ``{Challenges and Opportunities of mm-Wave Communication in 5G Networks},''
  {\em in Proc. IEEE International Conference on Cognitive Radio Oriented
  Wireless Networks and Communications (CROWNCOM), Oulu, Finland}, June 2014.

\bibitem{Andrews:2011bg}
J.~G. Andrews, F.~Baccelli, and R.~K. Ganti, ``{A Tractable Approach to
  Coverage and Rate in Cellular Networks},'' {\em IEEE Transactions on
  Communications}, vol.~59, no.~11, pp.~3122--3134, 2011.

\bibitem{Jiang:08}
T.~Jiang and Y.~Wu, ``{An Overview: Peak-to-Average Power Ratio Reduction
  Techniques for OFDM Signals},'' {\em IEEE Transactions on Broadcasting},
  vol.~54, no.~2, pp.~257--268, 2008.

\bibitem{SundeepADCmmW:15}
O.~Orhan, E.~Erkip, and S.~Rangan, ``{Low Power Analog-to-Digital Conversion in
  Millimeter Wave Systems: Impact of Resolution and Bandwidth on
  Performance},'' {\em available at: http://arxiv.org/abs/1502.01980v1.}

\bibitem{Decouple:14}
H.~Elshaer, F.~Boccardi, M.~Dohler, and R.~Irmer, ``{Downlink and Uplink
  Decoupling: a Distruptive Architectural Design for 5G Networks},'' {\em in
  Proc. IEEE Global Communications Conference (GLOBECOM), Austin, TX, United
  Sates}, 2014.

\bibitem{BoccardiDUde:15}
F.~Boccardi, J.~G. Andrews, H.~Elshaer, M.~Dohler, S.~Parkvall, P.~Popovski,
  and S.~Singh, ``{Why to Decouple the Uplink and Downlink in Cellular Networks
  and How to Do It},'' {\em available at: http://arxiv.org/abs/1503.06746.}

\bibitem{JHParkGeo:15}
J.~Park, {\em Building statistics extraction via QGIS from geographical data in
  Seoul, Korea, available at: http://tiny.cc/jhparkgeo}.

\bibitem{QGIS}
{QGIS,} {\em available at: http://www.qgis.org.}

\bibitem{Rel12:2015}
{3GPP TS 36.306, Evolved Universal Terrestrial Radio Access (E-UTRA) User
  Equipment (UE) Radio Access Capabilities.}

\bibitem{WRoh:14}
W.~Roh, S.~Ji-Yun, J.~Park, L.~Byunghwan, J.~Lee, K.~Yungsoo, J.~Cho, K.~Cheun,
  and F.~Aryanfar, ``{Millimeter-Wave Beamforming as an Enabling Technology for
  5G Cellular Communications: Theoretical Reasibility and Prototype Results},''
  {\em IEEE Communications Magazine}, vol.~52, no.~2, 2014.

\bibitem{AkiraDAC:13}
K.~Okada, K.~Kondou, M.~Miyahara, M.~Shinagawa, H.~Asada, R.~Minami,
  T.~Yamaguchi, Y.~Hino, T.~Sato, H.~Sakaguchi, N.~Shimasaki, T.~Ito,
  Y.~Takeuchi, N.~Li, Q.~Bu, R.~Murakami, K.~Bunsen, K.~Matushita, M.~Noda, and
  A.~Matsuzawa, ``{Full Four-Channel 6.3-Gb/s 60-GHz CMOS Transceiver with
  Low-Power Analog and Digital Baseband Circuitry},'' {\em IEEE Journal of
  Solid-State Circuits}, vol.~48, no.~1, pp.~46--65, 2013.

\bibitem{Ferenc:07}
J.-S. Ferenc and Z.~N{\'e}da, ``On the size-distribution of poisson voronoi
  cells,'' {\em Physica A: Statistical Mechanics and Its Applications},
  vol.~35, no.~2, pp.~518--526, 2007.

\bibitem{HaenggiSG}
M.~Haenggi, {\em {Stochastic Geometry for Wireless Networks}}.
\newblock Cambridge University Press, 2013.

\bibitem{AbramowitzBook:HandbookFunctions:1995}
M.~Abramowitz and I.~Stegun, {\em {Handbook of Mathematical Functions}}.
\newblock Dover, 1972.

\end{thebibliography}

\end{document}